\RequirePackage{fix-cm}
\documentclass[smallextended]{svjour3}       
\smartqed  

\usepackage{booktabs}
\usepackage{graphicx}
\usepackage{url}
\usepackage{amsmath,amssymb,amsfonts}
\usepackage{algorithmic}
\usepackage{textcomp}
\usepackage[table]{xcolor}
\usepackage{booktabs}
\usepackage{tabularx}
\usepackage{import}
\usepackage{booktabs}
\usepackage{soul}
\usepackage{tcolorbox}
\usepackage[normalem]{ulem}
\colorlet{Mycolor1}{green!10!orange!90!}
\usepackage{enumerate}
\usepackage{listings}
\AtBeginDocument{\DeclareCaptionSubType{lstlisting}}
\usepackage{caption,subcaption}
\usepackage{multirow}
\usepackage[numbers]{natbib}
\usepackage{latexsym}

\definecolor{codegreen}{rgb}{0,0.6,0}
\definecolor{codegray}{rgb}{0.5,0.5,0.5}
\definecolor{codepurple}{rgb}{0.58,0,0.82}
\definecolor{backcolour}{rgb}{0.95,0.95,0.92}

\lstdefinestyle{mystyle}{
    backgroundcolor=\color{backcolour},   
    commentstyle=\color{codegreen},
    keywordstyle=\color{magenta},
    numberstyle=\tiny\color{codegray},
    stringstyle=\color{codepurple},
    basicstyle=\ttfamily\footnotesize,
    breakatwhitespace=false,         
    breaklines=true,                 
    captionpos=b,                    
    keepspaces=true,                 
    numbers=left,                    
    numbersep=5pt,                  
    showspaces=false,                
    showstringspaces=false,
    showtabs=false,                  
    tabsize=2
}

\def\BibTeX{{\rm B\kern-.05em{\sc i\kern-.025em b}\kern-.08em
    T\kern-.1667em\lower.7ex\hbox{E}\kern-.125emX}}

\newcommand{\heng}[1]{\textcolor{red}{{\it [Heng says: #1]}}}
\newcommand{\Foutse}[1]{\textcolor{blue}{{\it [Foutse says: #1]}}}

\begin{document}

\title{Bug Characteristics in  Quantum Software Ecosystem}

\author{Mohamed Raed El Aoun      \and
        Heng Li     \and
        Foutse Khomh     \and
        Lionel Tidjon}

\authorrunning{Mohamed Raed El Aoun \and Heng Li \and Foutse Khomh \and Lionel Tidjon}


\institute{Mohamed Raed El Aoun, Heng Li, Foutse Khomh, Lionel Tidjon \at
              Department of Computer Engineering and Software Engineering \\
              Polytechnique Montreal \\
              Montreal, QC, Canada \\
              \email{\{mohamed-raed.el-aoun, heng.li, foutse.khomh, lionel.tidjon\}@polymtl.ca}
}

\date{Received: date / Accepted: date}

\maketitle

\begin{abstract}
With the advance in quantum computing (e.g., IBM's quantum computers) in recent years, quantum software becomes vital for exploring the full potential of quantum computing systems.
Quantum programming is different from classical programming in many different ways, for example, the state of a quantum program is probabilistic in nature, and a quantum computer is error-prone due to the instability of quantum mechanisms. Therefore, the characteristics of bugs in quantum software projects may be very different from that of classical software projects.  
This work aims to understand the characteristics of bugs occurring in quantum software projects, in order to provide insights that can help devise effective testing and debugging mechanisms for quantum software projects.  
 To achieve this goal, we conduct an empirical study on the bug reports (in the forms of pull requests and issue reports) of 125 quantum software projects hosted on GitHub. 
 We observe that quantum software projects are more buggy than comparable classical software projects and that quantum project bugs are more costly to fix (in terms of the code changed) than classical project bugs.

We also identify the types of these bugs and the quantum programming components (e.g., state preparation) where they occurred. Our study shows that the bugs are spread across different components, but quantum-specific bugs particularly appear in the compiler, gate operation, and state preparation components. The three most occurring types of bugs are Program anomaly bugs, Configuration bugs, and Data type and structure bugs. 
Our study highlights some particularly challenging areas in quantum software development that are different from that of traditional software development, such as the lack of scientific quantum computation libraries that implement comprehensive mathematical functions for quantum computing algorithms and quantum gate operation definitions. Quantum developers also seek specialized data manipulation (e.g, array manipulation) libraries dedicated to quantum software engineering such as \texttt{Numpy} for quantum computing. Our findings also provide insights for future work to advance the quantum program development, testing, and debugging of quantum software, such as providing tooling support for debugging low-level circuits. 
\end{abstract}

\keywords{
Quantum computing, quantum programming, quantum software engineering, issue reports, quantum program bug.
}

\vspace{-5pt}
\section{Introduction}
Quantum computing has achieved a series of important milestones in recent years. For example, D-Wave claimed the first commercial quantum computer in 2011~\cite{DWave2011}. Tech giants such as IBM, Amazon, Google, and Microsoft are racing to build their quantum computers. 
Quantum computers are expected to make revolutionary computation improvements over modern classical computers in certain areas, such as optimization, simulation, and machine learning \cite{mueck2017quantum, piattini2021toward}.

The rapid development of quantum computers has driven the development of quantum programming languages and quantum software~\cite{zhao2020quantum}, with many of them released as open source~\cite{qosf2021opensource}. A variety of quantum programming frameworks and languages have been introduced, such as Qiskit~\cite{Qiskit}, Cirq~\cite{cirq_developers_2021_4586899}, and Q\#~\cite{Qsharp}. 
IBM's Qiskit is a Python-based software development toolkit for developing quantum applications that can run on quantum simulators or real quantum computers (e.g., IBM Quantum Cloud). 

Quantum software, by its nature, is drastically different from classical software. For example, a classical software system is executed sequentially and the status of the system is typically deterministic. However, a quantum software system is intrinsically parallel 
and can have multiple possible states at the same time~\cite{moguel2020roadmap}. 
In addition, as quantum computers are error-prone due to the instability of quantum mechanisms, the output of a quantum software system is often noisy~\cite{nachman2020unfolding}. 
Thus, the bugs of quantum software may possess characteristics that are very different from those in classical software.

In this paper, we perform an empirical study on 125 open-source quantum software projects hosted on GitHub. These quantum software projects cover a variety of categories, such as quantum programming frameworks, quantum circuit simulators, or quantum algorithms. An analysis of the development activity of these selected projects show 
a level of development activities similar to that of classical projects hosted on GitHub. 
To understand the characteristics of bugs occurring in quantum software projects, we examined 
the following two research questions. 
\begin{description}
\item [\textbf{RQ1:}] \textit{How buggy are quantum software projects and how do developers address them?}
In this research question, we compare the distribution of bugs in quantum software projects and classical software projects, as well as developers' efforts in addressing these bugs.
We observe that quantum software projects are more buggy than comparable classical software projects. In addition, fixing quantum software bugs is more costly than fixing their classical counterparts.
Our results indicate the need for efforts to help developers identify quantum bugs in the early development phase (e.g., through static analysis), to reduce bug reports occurrence and bug fixing efforts.
\item [\textbf{RQ2:}] \textit{What are the characteristics of quantum software bugs?}
We qualitatively studied a statistically representative sample of quantum software bugs to understand their characteristics. In particular, we analyzed the quantum software components (e.g., quantum measurement) where these bugs occurred and examined the nature of these bugs (e.g., performance bugs). We observed that both quantum computing-related bugs and classical bugs occur in quantum computing components. The gate operation component is the most buggy. We identified a total of 13 different types of bugs occurring in quantum components. The three most occurring types of bugs are  \texttt{Program anomaly bugs}, \texttt{Configuration bugs}, and \texttt{Data type and structure bugs}. 
These bugs are often caused by the wrong logical organization of the quantum circuit, state preparation, gate operation, measurement, and state probability expectation computation.


\end{description}

Our work is important to guide future works that aim to develop methodology and tooling to support the identification and diagnosis of quantum software bugs. Some of our bug results emphasize a quantum-specific approach to identifying bugs not detected by traditional techniques. Our findings can be beneficial to quantum computing developers. They can learn from the most occurring bug types and avoid them in future work. Our results on which component is more buggy can help guide developers' testing efforts. As quantum computing frameworks are still in their early days and not yet widely used, our study of bug types can have a beneficial impact on the future releases of the quantum frameworks.

Because quantum computing is still in its early stages, there have been very few empirical studies on quantum computing bugs. In 2021, Campos and Souto~\cite{Campos2021QBugsAC} highlighted the need to study quantum computing bugs. In the same year, Zhao et al~\cite{abs-2103-09069} studied 36 bugs from a single quantum computing framework (Qiskit). 
In 2022, Matteo and Michael~\cite{abs-2110-14560} identified the patterns of 223 real-world bugs in 18 quantum software projects. In this paper, we study a larger set of bugs from a larger number of projects and perform a deeper analysis of the characteristics and types of quantum bugs.

In essence, this work makes the following contributions:
\begin{itemize}
    \item A preliminary study of the status and a categorization of quantum software projects. 
    \item A quantitative study of the bugs of 125 quantum software projects with comparison to bugs of classical software projects.
    \item An in-depth qualitative study of 333 real-world bugs in 125 quantum software projects.
    \item An investigation of the types of quantum bugs, and their distribution across the quantum components.
    \item Insights about the challenges and the complexity to fix the bugs.
\end{itemize}
\noindent \textbf{Paper organization.} The rest of the paper is organized as follows. In Section~\ref{sec:background}, we discuss key concepts of 
quantum software engineering. 
In Section~\ref{sec:setup}, we introduce the experimental setup of our study.
A preliminary study on the characteristics of quantum software projects is presented in Section~\ref{sec:prestudy}.
In Section~\ref{sec:results}, we present the answers to our research questions. 
Section~\ref{sec:relatedwork} reviews the related literature, while Section~\ref{sec:threats} discusses threats to the validity of our findings. Finally, Section~\ref{sec:conclusions} concludes the paper and discusses future works.

\section{Background} \label{sec:background}
In this section, we provide background knowledge about 
quantum computing, quantum software development, and the current quantum computing ecosystem. 

\subsection{Quantum Computing} \label{sec:background-qc}

While classical computers use bits in the form of electrical pulses to represent 1s and 0s, quantum computers use quantum bits (or qubits) in the form of subatomic particles such as electrons or photons to represent 1s and 0s. 
A qubit, unlike a classical bit, can be 0 or 1 with a certain probability, which is known as the \textit{superposition} principle~\cite{kaye2007introduction}. In simple words, a qubit can have two states at the same time.
Thus, a quantum computer consisting of multiple qubits is in many different states at the same time.
When a qubit is \textit{measured}, it collapses into a deterministic classical state.
 The status of two or more qubits can be correlated (or entangled) in the sense that changing the status of one qubit will change the status of the other(s) in a predictable way, which is known as the \textit{entanglement} phenomenon~\cite{kaye2007introduction}.  
The \textit{superposition} and \textit{entanglement} phenomenons give quantum computers advantages over classical computers in performing large-scale parallel computation~\cite{kaye2007introduction}.

\begin{figure}[htbp]
\vspace{-3mm}
\centerline{\includegraphics[width=0.7\textwidth]{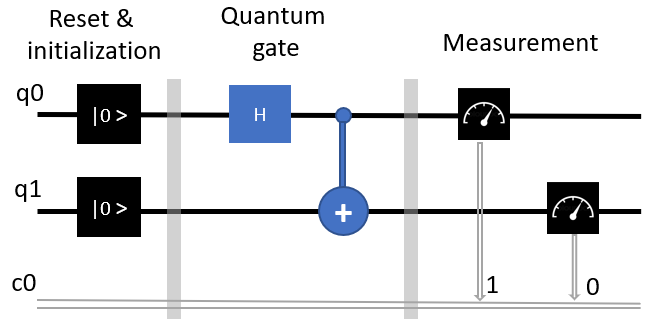}}
\vspace{-2mm}
\caption{An example quantum circuit.}
\label{fig:circuit}
\vspace{-8mm}
\end{figure}

Similar to classical logic gates (e.g., \texttt{AND}, \texttt{OR}, \texttt{NOT}), \textit{quantum logic gates} (or \textit{quantum gates}) alter the states (the probability of being 0 or 1) of the input qubits.
Like classical digit circuits, \textit{quantum circuits} are collections of quantum logic gates interconnected by quantum wires. 
Figure~\ref{fig:circuit} illustrates an example quantum circuit with two qubits (\textit{q0} and \textit{q1}). Below, we describe the three parts of the circuit:
\begin{enumerate}
    \item \emph{Reset and initialization}: The states of the two qubits are initialized as 0s. 
    \item \emph{Quantum gate}: A Hadamard (\textbf{\textit{H}}) gate denoted by a blue square is applied on the qubit \textit{q0}. The gate \textbf{\textit{H}} 
    generates a superposition state with equal probabilities for the states of 1 and 0.
    Then, a controlled-NOT \textbf{\textit{CX}} gate represented by a blue circle is applied on qubits \textit{q0} and \textit{q1}: the state of qubit \textit{q1} is flipped if and only if the state of qubit \textit{q0} is 1. Thus, the \textit{CX} gate creates entanglement between the pair of qubits \textit{q0} and \textit{q1} (i.e., the state of one qubit is predictable from the state of the other). 
    \item \emph{Measurement}: The measurement collapses the state of a qubit from a superposition state into a deterministic single state. 
    The output of this step is a qubit with the most probable state. 
\end{enumerate}

\subsection{Quantum Software Development}

\begin{figure}[htbp]
\centerline{\includegraphics[width=0.85\textwidth]{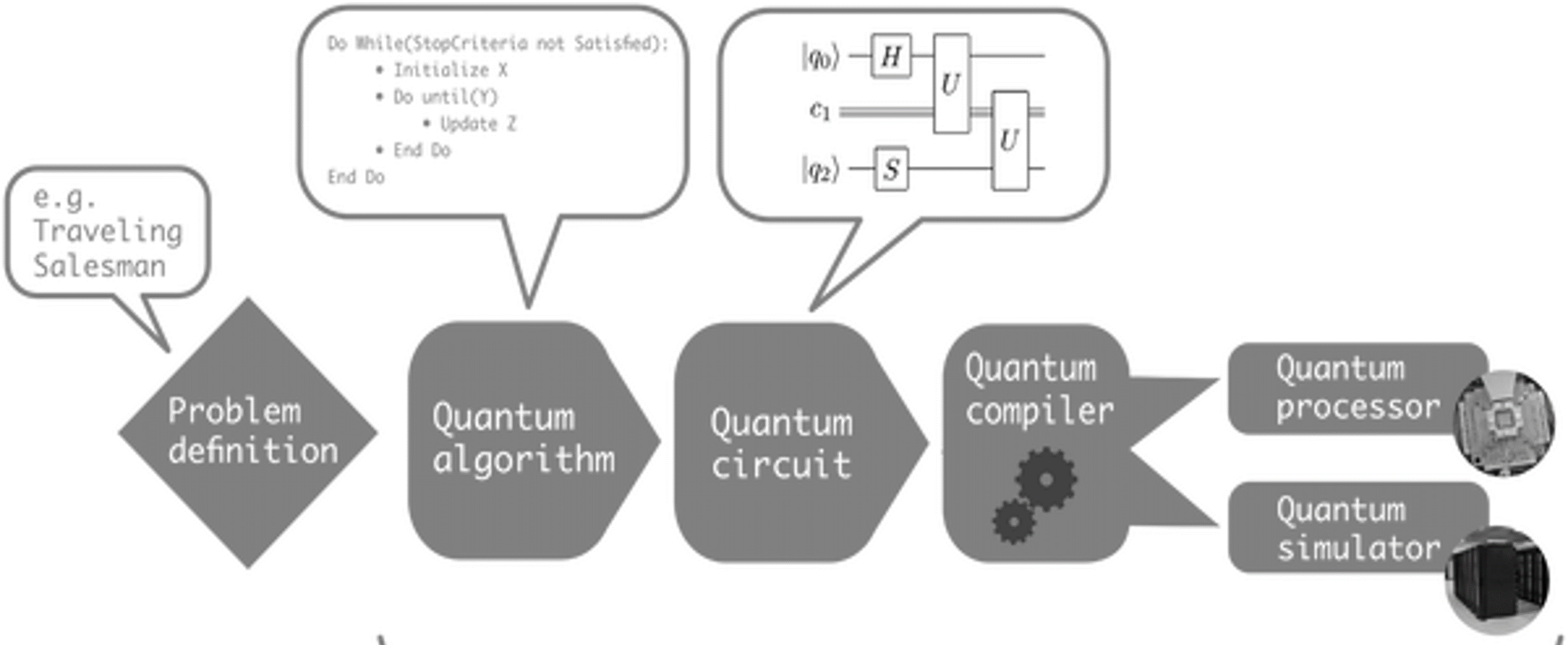}}
\caption{Quantum software workflow on a gate-model quantum computer \cite{Q_2018}}.
\label{fig:gate_model}
\end{figure}

Quantum software development typically follows the \texttt{quantum gate-model} to accomplish a certain task~\cite{Q_2018}. In this model, the problem is expressed in terms of quantum gates (described in Section \ref{sec:background-qc}). 
In Figure~\ref{fig:gate_model}, we illustrate an example of a quantum algorithm workflow using the gate model. The first step is to define the problem. In this example, we define the Travelling Salesman Problem (TSP). Secondly, based on the nature of the problem we need to choose the most suitable algorithm to find a solution. In our example, we can consider the quantum approximation algorithm \cite{farhi2014quantum}. This algorithm was proposed to find the optimal solutions using the gate model. Next, the quantum algorithm has to be implemented in quantum code which is then compiled into a quantum circuit consisting of quantum gates. 
Finally, the quantum circuit will be executed on a quantum computer or a simulator running on a classical computer.

\begin{lstlisting}[language=Python, caption=Quantum code example based on Qiskit that produces the quantum circuit shown in Figure \ref{fig:circuit},label={lst:code}]
import qiskit as q
# create register to store bits
qr = q.QuantumRegister(2)
cr = q.ClassicalRegister(2)
#create the circuit
circuit = q.QuantumCircuit(qr, cr)
#0th index on the quantum register 
circuit.h(qr[0])
#apply CX gate (control_bit, target_bit)
circuit.cx(qr[0], qr[1])
# measure quantum bit into cassical bit
circuit.measure(qr, cr)
\end{lstlisting}

Listing~\ref{lst:code} shows a code snippet based on the Qiskit python library that produces the quantum circuit shown in Figure~\ref{fig:circuit}. First, we import the 
\textit{Qiskit} library (line 1). Then, lines 3 and 4 create 2 qubits and 2 classical bits which form a quantum circuit in line 6. Line 8 applies the Hadamard (\textit{H}) gate on the first qubit, which generates a superposition of the qubit with equal probabilities of being 1 and 0. Then, line 10 applies a controlled-NOT (\textit{CX}) gate on the output of the Hadamard gate and the second qubit, which generates entanglement between the two qubits. Finally, line 12 measures the final states of the two qubits and maps the measurement results to the two classical bits.

\subsection{Quantum computing ecosystem}
\begin{figure}[htbp]
\centerline{\includegraphics[width=\textwidth]{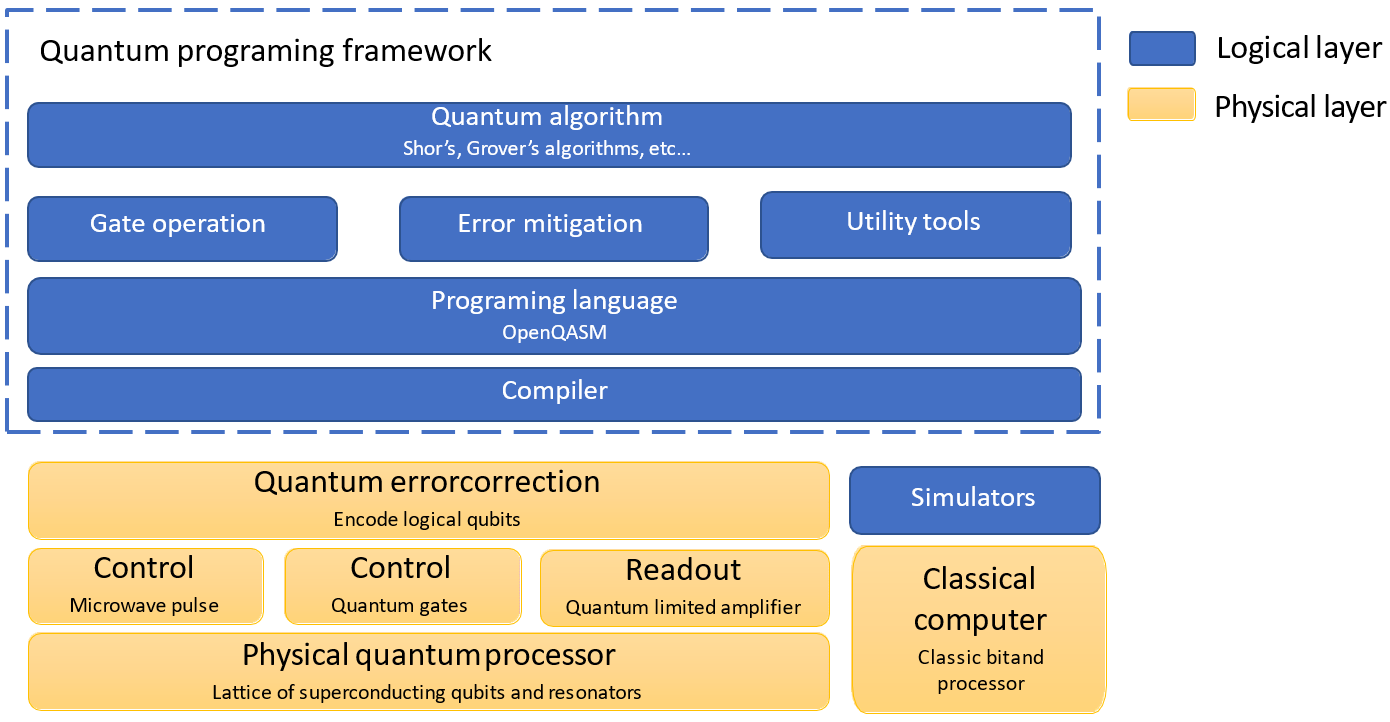}}
\caption{Quantum computing ecosystem overview.}
\label{fig:ecosystem}
\end{figure}
Quantum computing aims to solve problems that are challenging for a classical computer using principles of quantum mechanics. The Quantum computing ecosystem is starting to take shape, coalescing around hardware (quantum computers), software such as quantum programming frameworks, quantum programming languages, utility tools, and libraries (i.e., testing, error mitigation). In Figure~\ref{fig:ecosystem}, we present an overview of the state of the quantum ecosystem nowadays. This overview was inspired by Qiskit~\cite{Qiskit}, IBM-Q~\cite{IBM_q}, and an online article~\cite{Ecosystem_evo}.
From Figure~\ref{fig:ecosystem}, we can see that the quantum ecosystem has two different layers. A physical layer that includes both quantum and classical hardware, as well as a logical layer that covers the quantum computing software. 
The physical layer contains:
\begin{itemize}
    \item \textbf{Physical quantum processor:} A quantum circuit on a chip with a  size of hundreds of nanometers of quantum elements such as atoms and molecules known as qubit~\cite{phdthesis}.
    \item \textbf{Microwave pulse:} Device used to generate pulses to control and measure qubits fabricated on superconducting circuits~\cite{gao2020pulsequbit}.
    \item \textbf{Quantum gates:} Physical quantum gates and building blocks of quantum circuits~\cite{brandl2017quantum}
    \item \textbf{Quantum limited amplifier:} Amplification of the quantum signals (pulse) while adding the minimum amount of noise tolerated by quantum mechanics~\cite{Roy_2018}.
    \item \textbf{Quantum error correction:} Encode the logical qubits into multiple physical qubits while protecting the quantum states and actively correcting the errors~\cite{Cramer}.
    \item \textbf{Classical computer:} Traditional computer stores information in classical bits that are represented logically by either a 0 (off) or a 1 (on) \cite{ClassicalComputer}.  
    
\end{itemize}
The logical layer contains:
\begin{itemize}
    \item \textbf{Simulator:} Libraries and software to simulate the quantum computer behavior on a classical computer.
    \item \textbf{Compiler:} Tools and software used to compile and optimize the quantum circuits.
    \item \textbf{Programming language:} Quantum programming languages (e.g., Q\#~\cite{Qsharp}) are implemented in development kits to support the development of quantum algorithms.
    \item \textbf{Gate operation:} A set of unitary operations that are used to control the state of qubits.
    \item \textbf{Error mitigation:} Libraries and algorithms for software-based quantum error correction.
    \item \textbf{Utility tools:} Libraries that are used to support quantum software development activities, such as testing and debugging.
    \item \textbf{Quantum algorithms:} A collection of quantum algorithms that runs on top of a quantum computer or a simulator. 
\end{itemize}

\section{Study Definition and Data Extraction Methodology} \label{sec:setup}
\begin{figure}[htbp]
\vspace{-5mm}
\centerline{\includegraphics[width=1.0\textwidth]{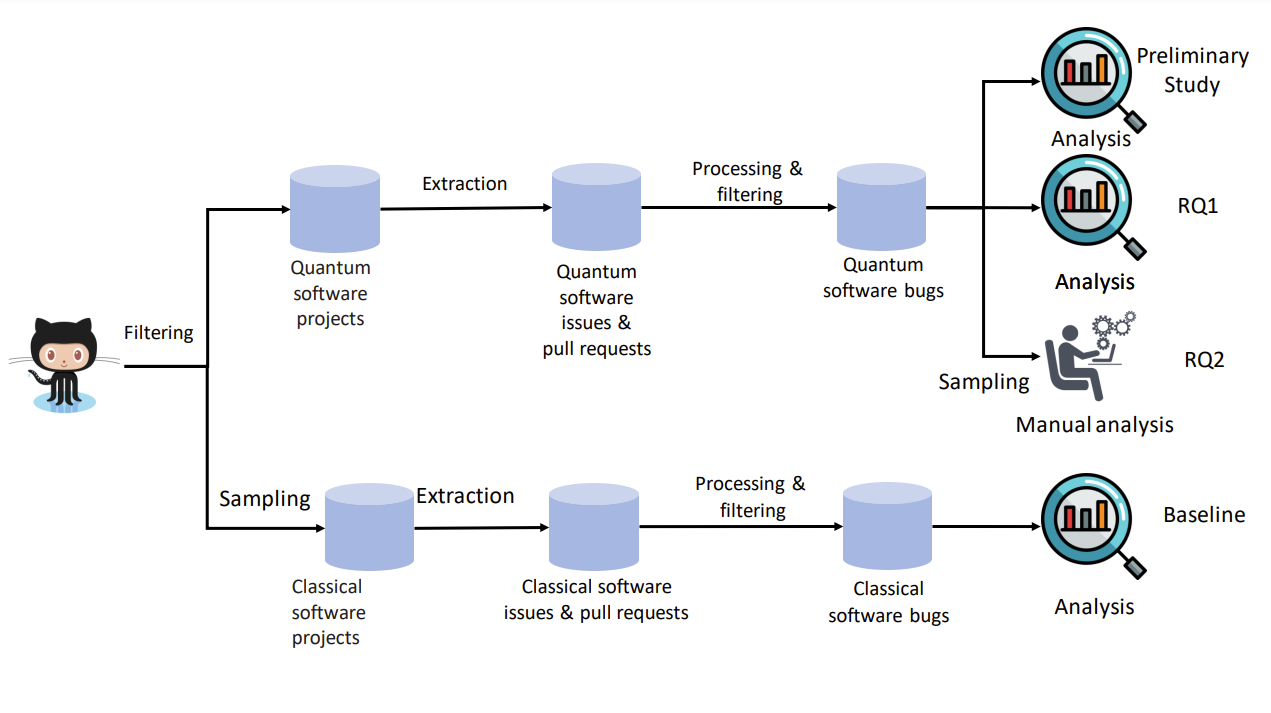}}
\caption{Overview of our study.}
\label{fig:overview}
\end{figure}

The \emph{goal} of this study is to understand the characteristics of bugs occurring in quantum software projects. The \emph{perspective} is that of researchers and tool builders interested in developing methodologies and tools to support the identification and diagnosis of quantum software bugs. The \emph{context} of the study is 125 open-source quantum software projects hosted on GitHub. To achieve the study's goal, we proceed in three steps. First, in a preliminary study, we examine the nature of quantum projects, comparing them to classical projects. This preliminary study is important to identify the specific characteristics of quantum projects that could affect the bugs. Next, we conduct a quantitative study, comparing the distribution of bugs in quantum software projects and classical software projects, as well as developers’ efforts in addressing these bugs. Finally, we qualitatively
studied a statistically representative sample of quantum software bugs and build a taxonomy of quantum software bugs. 
Figure~\ref{fig:overview} depicts the methodology of our study. First, we collect quantum and classical projects from GitHub. Then, we collect the pull requests and issues of these projects using GitHub issues API. Next, we apply a set of heuristics to identify quantum software project bugs and classical project bugs. Finally, we perform open coding and apply statistical analysis to answer our research questions. The following sections elaborate in detail on each of these steps. 
\vspace{-10mm}
\subsection{Quantum software projects collection}
We followed two steps to collect quantum software projects. 

\noindent\textbf{Step 1: Searching candidate projects.} To select representative projects from the quantum ecosystem, we searched through the GitHub \texttt{search API}. We used two ways to search for quantum software projects. 

\begin{itemize}
    \item First, we conduct a \texttt{keyword search} using the GitHub search API. To identify the quantum software projects in GitHub, we reduce our search scope and select the projects with a description or a topic that contains the word ``quantum computing'' (the word quantum computing is case sensitive to not miss relevant projects). 
    
    \item Next, we \texttt{search for code} that imported quantum programming libraries. Our code search items \textbf{$Q_{code}$} contain import statements from popular quantum computing libraries such as \texttt{Qiskit} and \texttt{Cirq} developed by pioneers like Google and IBM. Using the \texttt{search code} feature, we search for \textbf{$Q_{code}$} in the quantum projects source code to identify the quantum software projects. Our searched items \textbf{$Q_{code}$} include: \texttt{import qiskit, import cirq, from cirq import, from qiskit import, import tensorflo-\\w\_quantum, from tensorflow\_quantum import, from braket.circuits\\ import, import braket.circuits}. 
\end{itemize}

In addition, we limit our search results based on the following criteria:
\begin{itemize}
    \item To allow us to better understand the issues and content of the projects, the description of the project must be in English. 
    \item Since we are looking to characterize the bugs in the quantum software projects, the projects must have issues. 
\end{itemize}

At the end of this step, we obtained \textbf{$Q_{P-initial}$} containing a total of 2,105 unique quantum project.
\begin{figure}
\centering
\captionsetup{justification=centering}
\begin{minipage}{.5\textwidth}
  \centering
  \includegraphics[scale=0.17]{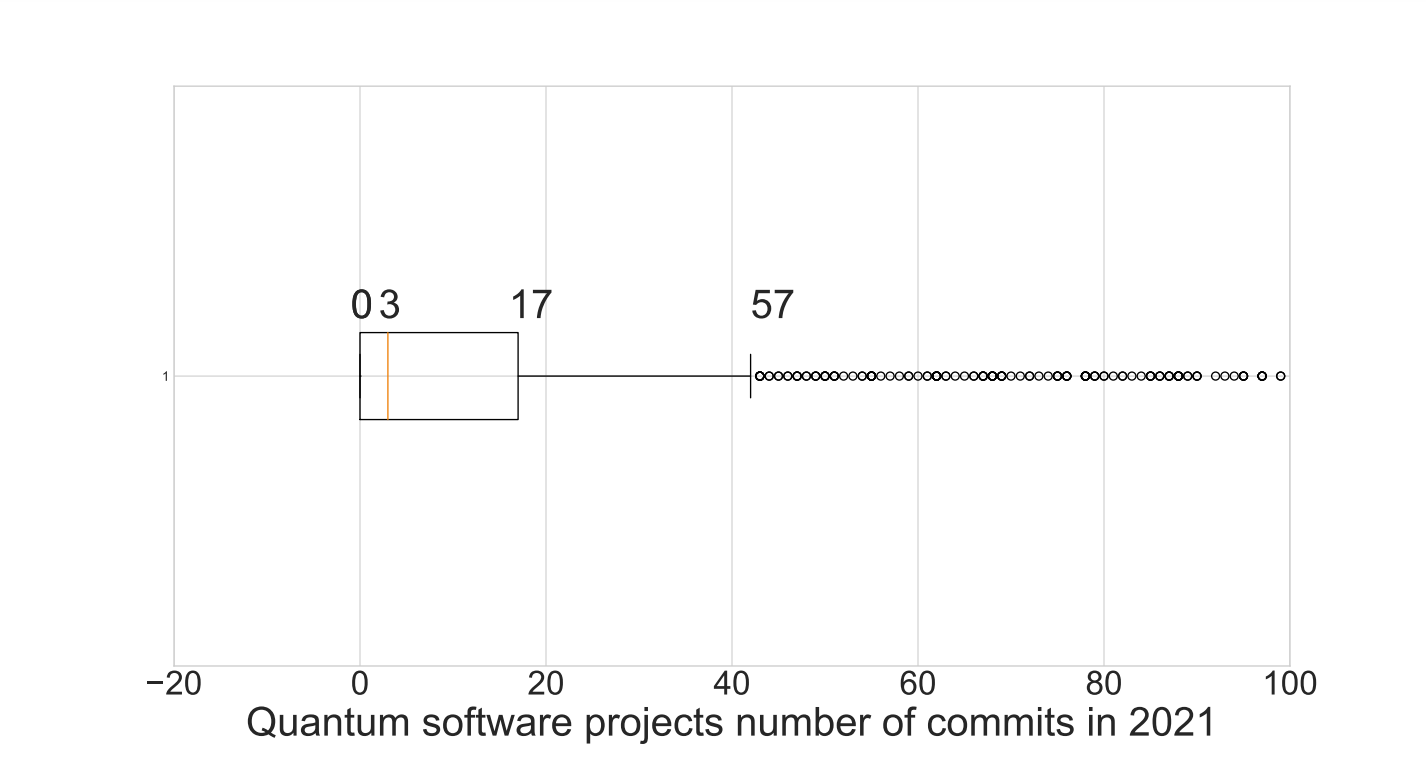}
  \captionof{figure}{Distribution of the number of commits in quantum software projects in the past year}
  \label{figure:15_commit}
\end{minipage}%
\begin{minipage}{.5\textwidth}
  \centering
  \includegraphics[scale=0.17]{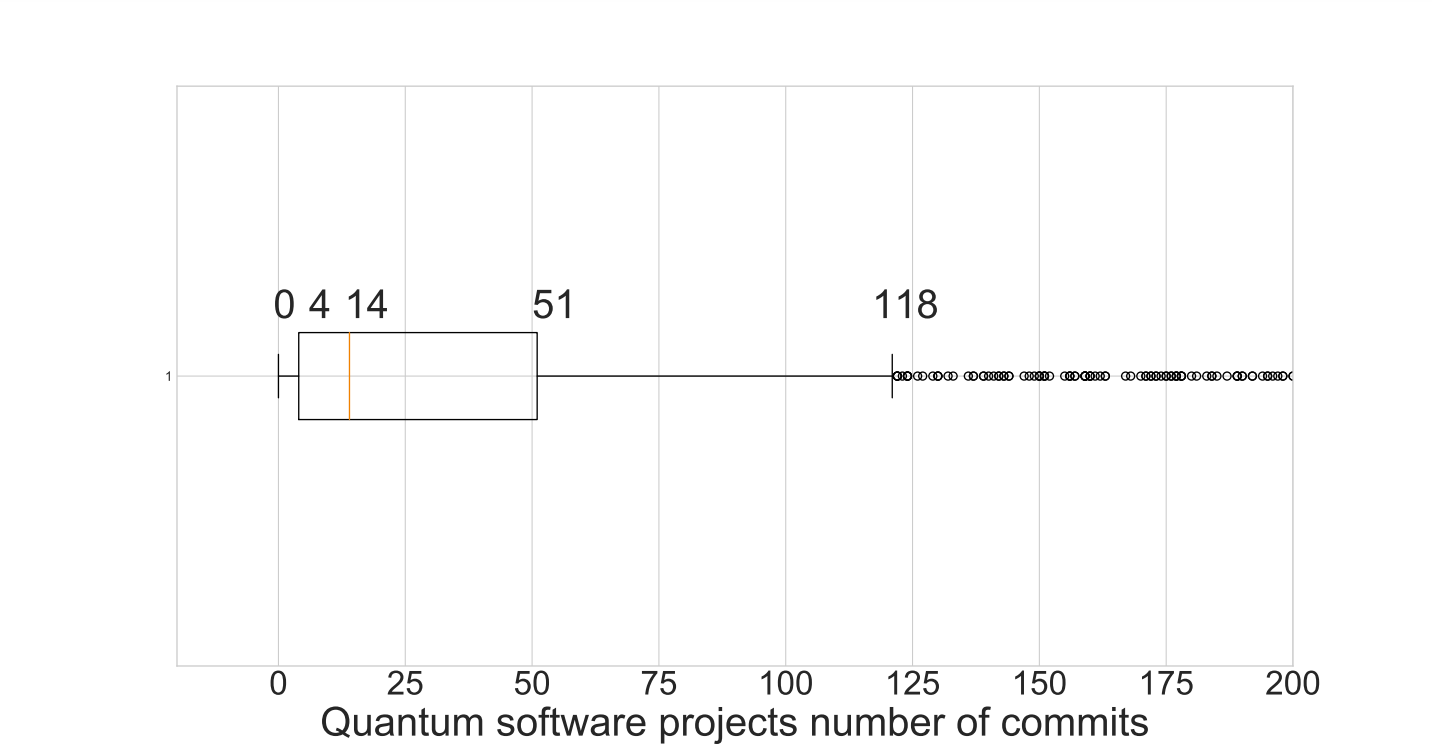}
  \captionof{figure}{Distribution of the number of commits in quantum software projects}
  \label{figure:100_commit}
\end{minipage}
\vspace{-3mm}
\end{figure}

\noindent\textbf{Step 2: Filtering collected projects.} 
We filter \textbf{$Q_{P-initial}$}, the collected set of quantum software projects following guidelines proposed by Kalliamvakou et al~\cite{Kalliamvakou2015AnIS}. Specifically, we selected projects based on their \texttt{number of commits in 2021}, their \texttt{ total number of commits}, and their \texttt{number of contributors}. These 3 criteria ensure that we identify the most active projects, remove the abandoned projects, and filter out the quantum computing projects related to documentation, lecture notes, and student assignments~\cite{Kalliamvakou2015AnIS}. 
\begin{enumerate}
    \item We keep projects with at least 51 commits to ensure that they have sufficient development activity 
    and to avoid student assignment~\cite{Businge2018CloneBasedVM}~\cite{Businge2019StudyingAA}. Figure \ref{figure:100_commit} show the distribution of the \texttt{total number of commits} in \textbf{$Q_{P-initial}$}. Since the third quantile is 51, we used it as our threshold to filter the set of projects \textbf{$Q_{P-initial}$}.
    \item We select the projects with at least 17 commits in the past year (i.e., 2021) to remove the abandoned and inactive projects. Figure \ref{figure:15_commit} shows the distribution of the number of commits in the past years in \textbf{$Q_{P-initial}$}. We observe that 75\% of the projects have less than 17 commits in the past year, therefore we use this value as a threshold to filter the set of projects \textbf{$Q_{P-initial}$}.
    \item We retain projects with more than one contributor to avoid selecting toy projects. 
\end{enumerate}

After applying these filtering criteria, we manually inspected the remaining projects' descriptions and removed non-quantum projects (e.g,, projects related to quantum physics, documentation, or lecture notes).

Finally, we ended up with a total of 125 projects (\textbf{$Q_{P-final}$}) that are directly related to quantum computing.

\subsection{Classical projects collection} 
To understand the quantum software projects' characteristics, we need a baseline for our analysis. Hence, we collected a set of classical software projects comparable to the quantum software projects. From GitHub, we selected classical projects that have the range of \texttt{number of stars}, \texttt{programming language}, \texttt{number of watcher}, similar to our quantum repositories and the repository must have \texttt{issues}. 
As the number of projects in the extracted list of classical software projects is very large, to have a representative baseline, we collected a random sample of 324 projects using a 95\% confidence interval level and a 5\% margin of error. 

\subsection{GitHub issues and pull requests collection}\label{sec:issue-extraction}
To identify the bugs and to uncover the cost to fix a bug in the quantum software projects, we collected the pull requests and issue reports for each of the studied projects, using GitHub issues API which considers every pull request as an issue. We collected the following fields: \texttt{issue number, state, title, repository url, issue url, pull request url, body, creation date, close date, and merge date}. Using GitHub issue API \cite{githubrestapi} for each project, we extracted \textbf{$B_{initial}$} a total of 59,249 issues and pull requests from \textbf{$Q_{P-final}$}. In the rest of the paper, refer to the issue reports and the pull requests as issue reports or issues. 

\subsection{Identification of quantum software projects bugs} \label{sec:bug-extraction}
To identify the bugs in GitHub issue reports, we first map each issue with its corresponding commit. For each project, we collect the related commits of each issue report using the GitHub API. To identify the commit or list of commits of an issue report, we look into the list of events of the issue report. If an issue is referenced in a commit, the \texttt{commit url} will appear in the list of events with the tag \texttt{referenced}. Therefore we collect the list of \texttt{commit url} that appears in the issues event.
We collect the commits of the issues report for each project using the \texttt{GitHub commit API} and collect the following fields of each commit: \texttt{the commit message}, the\texttt{ URL}, the \texttt{number of added}, \texttt{deleted} or \texttt{changed} files, and the \texttt{contributor identification}. The next step is to classify if an issue in \textbf{$B_{initial}$} describes a bug or not. For this purpose, we proceed as follows:
\begin{enumerate}
    \item In GitHub, developers tag the issues using the \texttt{Label} field. To identify bugs, we select the issues that has the tag \texttt{Label=bug or defect} and flag the issues as a \texttt{bug}.
    \item Following previous work~\cite{Tan2013BugCI}, we define a set of keywords \textbf{$K_{initial}$} \textbf{fix, error, crash, wrong, bug, issue, fail, correct} to detect bugs.
    \item For each issue report b $\in{B_{initial}}$, if b's title or body or commit message contains at least a word k $\in{K_{initial}}$, b is flagged as a bug. We ended up with a set \textbf{$B_{1}$} of 18,084 bug reports at the end of this step .
    \item To extend the set \textbf{$K_{initial}$} and identify further potential bugs $\in{B_{initial}}$, we select a random sample with 95\% confidence interval level and 5\% marge of error of the issues not flagged as bugs, i.e.,  \textbf{$\in\overline{B_{1}}$}. 
    \item One author manually inspected the commits of the random sample of issue reports and defined an extended set of \textbf{$K_{initial}$}. Then, during a meeting, three authors discussed the validity of the new set of keywords. Finally, all authors agreed on the final set \textbf{$K_{final}$} \textbf{fix, error, crash, wrong, bug, issue, fail, correct, exception, log, inf, insufficientResource, broke, resolve, abort, leak.}
    \item For each issue reports b \textbf{$\in\overline{B_{1}}$}, if b's title, body or commit message contains at least a word k $\in{K_{final}}$, b is flagged as a bug. We obtained a set \textbf{$B_{final}$} containing 19,564 bug reports.
\end{enumerate}
As a result of the process, we ended up with a total of 19,564 bug reports \textbf{$B_{final}$} that are 
related to quantum computing.

\subsection{Classical projects bug extraction}

We follow the same processes as described in Section \ref{sec:issue-extraction} and Section \ref{sec:bug-extraction} to extract the bugs of the selected classical software projects. In the 324 classical projects, we had a total of 60,779 issue reports in which we identified 13,165 bugs.

\section{\textbf{Preliminary study: Characteristics of quantum software projects}} \label{sec:prestudy}
In order the understand quantum programming challenges, we first want to understand the nature of the quantum software projects. 
In particular, we want to understand the types of the projects and compare their characteristics (e.g., developer activities) with classical software projects.

\subsection{\textbf{Approach}}
\noindent \textbf{Categorization of quantum software projects.} To identify the categories of quantum projects in GitHub, we manually analyzed the descriptions and the documentation of all the projects. No prior taxonomy was used. 
For each project, we assigned one label. In case a project is associated with more than one categories, which occurred in a few cases, we chose the most representative one.
Two authors of the paper jointly performed the manual labeling. 
Each project is labeled by both authors. We describe our labeling process below.
\begin{enumerate}
    \item \textbf{Initial labeling.} Each author labels all the projects independently. 
    \item \textbf{Discussion.} To have a consistent labeling strategy, we scheduled a meeting after the initial labeling and reached a consensus on the set of labels. A third author of the paper is involved in the meeting.
    \item \textbf{Revising labels.} Each author updated their labeling results after the meeting.
    \item \textbf{Resolving disagreement.} We had a final meeting to resolve the disagreement in the labeling results and reached a consensus for the label of each project. For each mismatched label, the three authors discussed and resolved the conflict.
\end{enumerate}

\noindent \textbf{Analyzing the characteristics of quantum software projects.} 
We study the characteristics of the quantum software projects as follows:
\begin{enumerate}
  \item In order to understand the maturity of quantum software projects, we look into the \texttt{duration} (i.e., from the creation date of the project until the date of our data extraction), and to study the the activity in the quantum projects, we look into the \texttt{number of releases}, \texttt{number of commits}, and \texttt{number of contributors}. 
  We used GitHub API~\cite{githubrestapi} to collect these metrics for each studied project.
  
  \item In order to understand the profile of the developers of quantum software projects, for each project we retrieve the user name of the contributors, then for each contributor, with the help of the GitHub events API~\cite{githubrestapi}, we look into its created events since the contributor joined GitHub. Following GitHub documentation~\cite{githubrestapi}, we select the events that describe the activity of the developers. Specifically, we focus on the following events: \texttt{CommitCommentEvent}, \texttt{IssueEvent}, \texttt{IssueCommentEvent}, \texttt{PullRequestEvent}, and \texttt{PushEvent}.  
\end{enumerate}

Steps 1 and 2 are repeated on the sample of the classical software projects. The results of the classical software projects are used as our baseline.

\subsection{\textbf{Results}}
\begin{figure}
\centering
\captionsetup{justification=centering}
\begin{minipage}{.5\textwidth}
  \centering
  \includegraphics[width=1\linewidth]{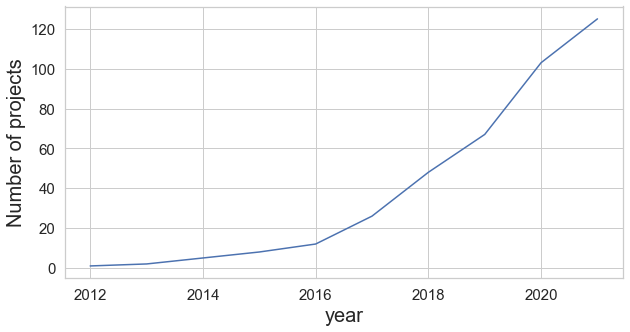}
  \captionof{figure}{Cumulative distribution function of the quantum software ecosystem projects over time (days)}
  \label{figure:CDF_nb_project}
\end{minipage}%
\begin{minipage}{.5\textwidth}
  \centering
  \includegraphics[width=.9\linewidth]{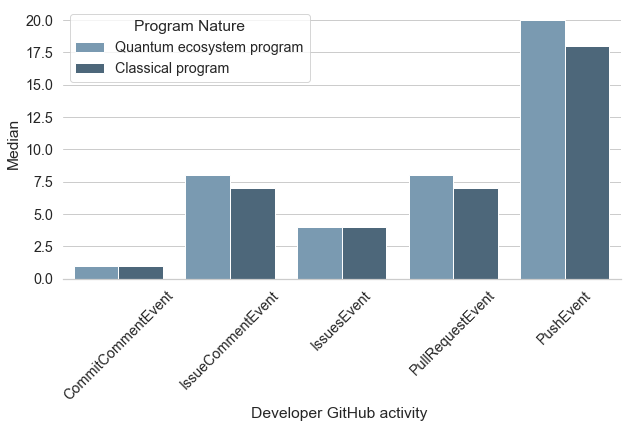}
  \vspace{-5mm}
  \captionof{figure}{Comparison of the GitHub activities of the developers  of classical and quantum projects 
  }
  \label{figure:dev}
\end{minipage}
\vspace{-2mm}
\end{figure}

\textbf{Quantum software projects have been steadily increasing in recent years.} Figure \ref{figure:CDF_nb_project} shows the number of quantum software projects over time.
We observe a significant increase in the number of projects after 2016. The increase may be explained by the rapid development of quantum computing technologies in recent years.
In 2017, IBM announced a working quantum computer with 50 qubits, that can maintain its quantum state for 60 microseconds ~\cite{IBM_raise}. Microsoft released its Q sharp programming language~\cite{Qsharp} in the same year. In 2018, IonQ released its first commercial trapped-ion quantum computer~\cite{IonQ}. In 2019, IBM released its first commercial quantum computer, the IBM Q System One \cite{IBM_one}. 

\begin{table*}[]
\tiny
\caption{Characteristics of the studied quantum projects.}
\begin{tabular}{lcccccc}
\hline
\multicolumn{1}{c}{\textbf{Program Category}} &
  \multicolumn{1}{l}{\textbf{\begin{tabular}[c]{@{}l@{}}nbr of\\  projects\end{tabular}}} &
  \textbf{\begin{tabular}[c]{@{}c@{}}\# of \\ releases\end{tabular}} &
  \textbf{\begin{tabular}[c]{@{}c@{}}\# of\\ commits\end{tabular}} &
  \textbf{\begin{tabular}[c]{@{}c@{}}\# of \\ contributors\end{tabular}} &
  \multicolumn{1}{l}{\textbf{LOC}} &
  \textbf{\begin{tabular}[c]{@{}c@{}}\# of\\languages\end{tabular}} \\ \hline
Quantum circuit simulators                                                & 26  & 6.5  & 113.0 & 5.0  & 19,145  & 9.0  \\ \hline
\begin{tabular}[c]{@{}l@{}}Quantum programming \\ framework\end{tabular}   & 22  & 10.0 & 163.5 & 24.0 & 49,981  & 12.0 \\ \hline
Quantum algorithms                                                        & 19  & 7.0  & 129.0 & 9.0  & 26,530  & 9.0  \\ \hline
Quantum utility tools                                                     & 19  & 7.0  & 126.0 & 6.0  & 27,830  & 9.0  \\ \hline
Quantum compilers                                                         & 9   & 3.0  & 179.0 & 10.0 & 175,364 & 16.0 \\ \hline
Quantum machine learning                                                  & 8   & 2.0  & 217.0 & 9.0  & 33,705  & 10.0 \\ \hline
Quantum circuit simulator                                                 & 7   & 19.0 & 29.0  & 15.0 & 87,316  & 13.0 \\ \hline
\begin{tabular}[c]{@{}l@{}}Experimental quantum \\ computing\end{tabular} & 7   & 10.0 & 147.0 & 15.0 & 66,088  & 13.0 \\ \hline
Quantum-based simulation                                                  & 2   & 16.5 & 459.0 & 84.5 & 92,819  & 14.5 \\ \hline
Quantum fun                                                               & 1   & 15.0 & 88.0  & 3.0  & 3,938   & 8.0  \\ \hline
\begin{tabular}[c]{@{}l@{}}\textbf{Quantum projects}\end{tabular} & 125 & 8.0  & 131.0 & 8.0  & 34,188.0 & 10   \\ \hline

\begin{tabular}[c]{@{}l@{}}\textbf{Classical projects}\end{tabular} & 324 & 12.0  & 268.0 & 2.0  & 15,357.0 & 7    \\ \hline
\multicolumn{6}{l}{\# all values are median values in the projects in that category}
\end{tabular}
\label{table:projectsSummary}
\vspace{-4mm}
\end{table*}

\textbf{We derived 9 categories of projects in the quantum computing software ecosystem covering quantum programming frameworks, simulators, tools, implementations of quantum algorithms, and applications such as machine learning.}
Table~\ref{table:projectsSummary} shows the characteristics of these quantum software projects by category. Below, we describe each category of projects.

\noindent\textbf{\texttt{Quantum circuit simulator}}: Libraries that are used to simulate quantum computation on a classical computer.

\noindent\textbf{\texttt{Quantum programming framework}}: A set of development kits, languages and libraries such as Q\#~\cite{Qsharp}, Qiskit~\cite{Qiskit} and pyEPR \footnote{\url{https://pyepr\-docs.readthedocs.io/en/latest/}} for quantum programming and quantum circuits design.

\noindent\textbf{\texttt{Quantum algorithms}}: Projects that provide one or more implemented quantum algorithms (i.e., grove\footnote{\url{https://github.com/Seeed\-Studio/grove.py}}).

\noindent\textbf{\texttt{Quantum machine learning}}: A set of libraries for implementing hybrid quantum-classical machine learning models. 
For example, TensorFlow Quantum (TFQ)\footnote{\url{https://www.tensorflow.org/quantum}} is a quantum machine learning library for prototyping hybrid quantum-classical ML models.

\noindent\textbf{\texttt{Quantum compilers}}: Tools used for the synthesis, compilation, and optimization of quantum circuits (i.e., QGL2 Compiler\footnote{\url{https://github.com/BBN\-Q/pyqgl2}}).

\noindent\textbf{\texttt{Quantum utility tools}}: Libraries and tools used to support quantum program development such as monitoring the load in quantum computers, API interface used to connect with quantum devices, or libraries dedicated for specific task development in the quantum programming workflow (i.e, state preparation, circuit visualization).

\noindent\textbf{\texttt{Experimental quantum computing}}: Libraries used for research and experimental calculation (i.e., artiq Compiler\footnote{\url{https://m-labs.hk/experiment-control/artiq/}}).

\noindent\textbf{\texttt{Quantum games}}: Games developed using quantum programming that run on simulators or quantum computers.

\noindent\textbf{\texttt{Quantum based-simulation}}: Libraries used for simulating quantum physics experiments on a quantum computer.

Table~\ref{table:projectsSummary} shows different metrics about the 125 studied projects and their distribution, along with the defined quantum ecosystem categories; including the number of releases, number of contributors, number of commits, number of lines of code (LOC), and number of programming languages used in the last release of the project. 

As shown in Table~\ref{table:projectsSummary}, the largest number of projects fall into the \texttt{quantum circuit simulators} category with 26 projects. As quantum computers are still not conveniently accessible to the public, simulators are widely used to execute quantum programs. Moreover, \texttt{quantum programming frameworks} is the second most frequent project category with 22 projects. As quantum computing is getting more popular, programming frameworks are starting to emerge. \texttt{Quantum algorithms } and \texttt{quantum utility tools} are the third most occurring categories with 19 projects for each. 
\texttt{\textbf{Quantum programming frameworks}} and \texttt{\textbf{utility tools}} are important to unlock the full potential of quantum computing systems. 
In Table~\ref{table:projectsSummary}, we observe that quantum software projects have a lower level of maturity (in terms of development activities) with a respective median number of commits and releases of 131 and 8, in comparison to classical projects which have a 268 median number of commits and 12 median number of releases.

\textbf{Quantum software project developers show a similar level of activities as that of classical software developers.}
In order to understand the activities of quantum project developers, in Figure~\ref{figure:dev} we compare the activities of quantum software projects and classical projects contributors by studying the developer's activities on GitHub since their registration in GitHub. Note that the quantum software projects and classical software projects are selected using the same criterion. We select five dimensions of activities from the GitHub \texttt{event API} : \texttt{commit on the commits}, \texttt{issues comments}, \texttt{created issues}, \texttt{created pull requests}, \texttt{push event} (i.e., commit and pull request). We chose these five dimensions because they describe how active a developer is in GitHub since joining. We observe some similarities between classical program developers and quantum program developers across the five dimensions. For example, quantum software developers and classical software developers have the same number of \texttt{issue event} with a median value of 3. However, quantum program developers are more active in other GitHub activities. From Figure \ref{figure:dev}, we observe that quantum developers have a slightly higher number of \texttt{issues comments event}, \texttt{pull request event}, and \texttt{push event}. These results show that developers of quantum software projects put more effort and face more issues.  
This observation motivated our investigation of the bug-proneness of 
quantum software projects. We also investigate how developers address  bugs in quantum programs, as well as the types and location of bugs in these programs. 

\vspace{-1mm}
\begin{tcolorbox}
\vspace{-1mm}
Quantum software projects become increasingly popular on GitHub. We observe a diverse range of quantum software projects, including quantum programming frameworks, tools, algorithms, and applications. We also observe that developers of quantum software projects are more active and face more issues in quantum programs. 
\vspace{-1mm}
\end{tcolorbox}
\vspace{-4mm}

\section{Characteristics of Bugs in Quantum Programs} \label{sec:results}
In this section, we report and discuss the results of our two research questions. For each research question, we present the motivation, the analysis approach, and the obtained results. 


\vspace{-3pt}
\subsection{How buggy are quantum software projects and how do developers address them?}
\subsubsection{\textbf{Motivation}}

In order to understand the characteristics of quantum software bugs, we first want to understand the extent to which quantum software projects are buggy, as well as how effectively developers address these bugs and their efforts in such activities.

\subsubsection{\textbf{Approach}}

Following the process described in Section~\ref{sec:setup}, we identified the bugs in the selected quantum software projects and classical software projects. To understand how developers address them, we further identify the fixed bugs, as well as the fixing time and the associated efforts.

\noindent \textbf{Identifying fixed bugs.}
A bug report is a GitHub issue report describing a bug. In GitHub, not all closed bug reports are solved. In some cases, a bug report can be closed while the problem persists. Thus, we had to identify the fixed and non-fixed bugs. We followed 3 steps to classify the bug reports.
\begin{enumerate}
    \item \textbf{Filtering the open issue reports.} Using the state field in the issues reports, we kept only the closed issues.
    \item \textbf{Identifying fixed bugs.} Following previous work~\cite{CATOLINO2019165} \cite{Tan2013BugCI}, we looked for the set of keywords \textbf{$F_{fix}$} in the GitHub issue report message and its corresponding commits: \textbf{\texttt{closes, close, closed, resolve, resolves, resolved, fix, fixes, fixed}}. To identify the fixed bugs, for each closed GitHub issue report, in the commit message, we look if it contains f $\in{F_{fix}}$. If the commit message contains at least one of the keywords we flag it as \texttt{Fixed}. 
    \item \textbf{Manual verification.} One author inspected a sample of 382 fixed bugs with a 95\% confidence level and 5\% marge of error. We found that our identification of fixed/non-fixed bugs has a precision of 1 and a recall equal to 0.97.
\end{enumerate}

\noindent \textbf{Bug fixing rate.} For each quantum or classical project, we calculate the bug fixing rate by dividing the number of fixed bugs by the total number of closed bugs (including fixed bugs and bugs closed without a fix) in that project. 

\noindent \textbf{Bug fixing duration.} For each fixed bug, we measure the bug fixing duration by calculating the difference between the bug fixing time and the bug creation time.

\noindent \textbf{Bug fixing effort.} To measure the effort involved in a bug fix, we collected the code changes of the commits that contributed to the bug fix. A change in a commit is an \texttt{added} or \texttt{deleted} code line. We also collected information about the number of changed files in the commits, which can be an \texttt{added}, \texttt{modified}, \texttt{renamed} or \texttt{deleted} file. For each bug fix, we measured two metrics:
\begin{itemize}
  \item \textbf{Number of changed files}: the \texttt{added}, \texttt{modified}, \texttt{renamed}, and \texttt{deleted} files in the commits that contribute to a bug fix.
  \item \textbf{Number of changed lines of code}: The number of \texttt{added} and \texttt{deleted} code lines in the commits that contribute to a bug fix.
\end{itemize}

\subsubsection{\textbf{Results}}

\textbf{Quantum software projects show more bugs (with a median number of 28 bugs) than classical software projects (with a median number of 13 bugs).} 
Table \ref{table:categories} provides a summary of the bugs in quantum software projects and the distribution of the bugs along with the project categories. A median of 28.0 bugs per project was detected in the studied quantum software projects, which is two times higher than the median number of bugs in the classical software projects. In light of these values, quantum programs are more buggy than classical programs. 
The higher number of bugs in a quantum project may be explained by the fact that quantum computers are error-prone and the output of quantum software is often noisy.
In particular, the quantum projects (e.g., projects in the categories of \textit{quantum-based simulation, experimental quantum computing, and quantum machine learning}) are the most buggy projects. This observation may be explained by the fact that these quantum projects categories rely more on quantum computing theories. Since there are no prior collection of mathematical algorithms for quantum theories and convenience functions that developers can leverage, the developers of quantum programs may be prioritizing the correct implementation of these complex theories over the quality of their code. 

The median value of the bug fix ratio in Table~\ref{table:categories} shows that 27.5\% of the quantum software bugs are closed but not fixed, compared to 21.1\% of classical software bugs which are closed without a fix. One author manually inspected a statistically representative sample of 382 closed-but-not-fixed bugs, representing a 95\% confidence level and 5\% confidence interval, to understand why these bugs are closed without a fix. We noticed that in some cases developers don't follow the best practices of GitHub and mention the issue number in the commit message with the fixing keywords, which makes detecting the bug fixes from commit messages difficult. As an example, we present issue number 5148 in Qiskit\footnote{\url{https://github.com/Qiskit/qiskit-terra/issues/5148}}. Besides, in some issue reports,  enhancements were labeled as bugs for future releases and closed by developers without notice in commit messages. For example, the issue id 20 in the qutip\footnote{\url{https://github.com/qutip/qutip/issues/20}}. Finally, we detected bugs that were closed because developers could not provide a fix, or  
it is difficult to reproduce the bug. For example the issue number 402 in \texttt{Cirq}\footnote{\url{https://github.com/quantumlib/Cirq/issues/402}}.

\begin{table}[]
\tiny 
\centering
\vspace{-2pt}
\caption{Distributions of bugs and bug fix ratio in quantum software projects}
\begin{tabular}{lccccc}
\hline
\multicolumn{1}{c}{\textbf{Project category}} &
  \textbf{\# of bugs} &
  \textbf{\begin{tabular}[c]{@{}c@{}}\# Bug fix \\ ratio\end{tabular}} &
  \textbf{\begin{tabular}[c]{@{}c@{}}Hours \\ to fix a bug\end{tabular}} &
  \textbf{\begin{tabular}[c]{@{}c@{}}\# file changed\\ in a commit\end{tabular}} &
  \textbf{\begin{tabular}[c]{@{}c@{}}\# LOC changed\\ in a commit\end{tabular}} \\ \hline
\begin{tabular}[c]{@{}l@{}}Quantum-based\\ simulation\end{tabular}       & 406.0 & 0.500 & 33.94 &  2.00 & 63.00 \\
\begin{tabular}[c]{@{}l@{}}Experimental quantum\\ computing\end{tabular} & 168.5 & 0.628 & 27.98 & 2.00 & 12.00 \\
\begin{tabular}[c]{@{}l@{}}Quantum Machine\\ learning\end{tabular}       & 104.0 & 0.736 & 66.05 &    2.00 & 38.0  \\
\begin{tabular}[c]{@{}l@{}}Quantum programming\\ framework\end{tabular}  & 88.0  & 0.513 & 34.60 & 2.00 & 30.00 \\
Quantum algorithms                                                       & 72.0  & 0.657 & 25.01 & 1.50 & 15.00 \\
Quantum compilers                                                        & 36.0  & 0.572 & 20.72 & 2.00 & 28.00 \\
\begin{tabular}[c]{@{}l@{}}Quantum circuit\\ simulators\end{tabular}     & 26.0  & 0.523 & 20.61 &  2.00 & 20.00 \\
Quantum utility tools                                                    & 21.0  & 0.689 & 15.40 & 2.00 & 11.00 \\ \hline
Quantum projects                                                         & 28.0  & 0.725 & 23.5  & 2.00 & 19.00  \\ \hline
Classical projects                                                       & 13.0  & 0.789 &   21.1    &  2.00 & 14.00 \\ \hline
\multicolumn{6}{l}{\# all values are median values in the projects in that category}
\end{tabular}
\label{table:categories}
\vspace{-3mm}
\end{table}

\begin{figure}
\centering
\captionsetup{justification=centering}
\begin{minipage}{.8\textwidth}
  \centering
  \includegraphics[width=1\linewidth]{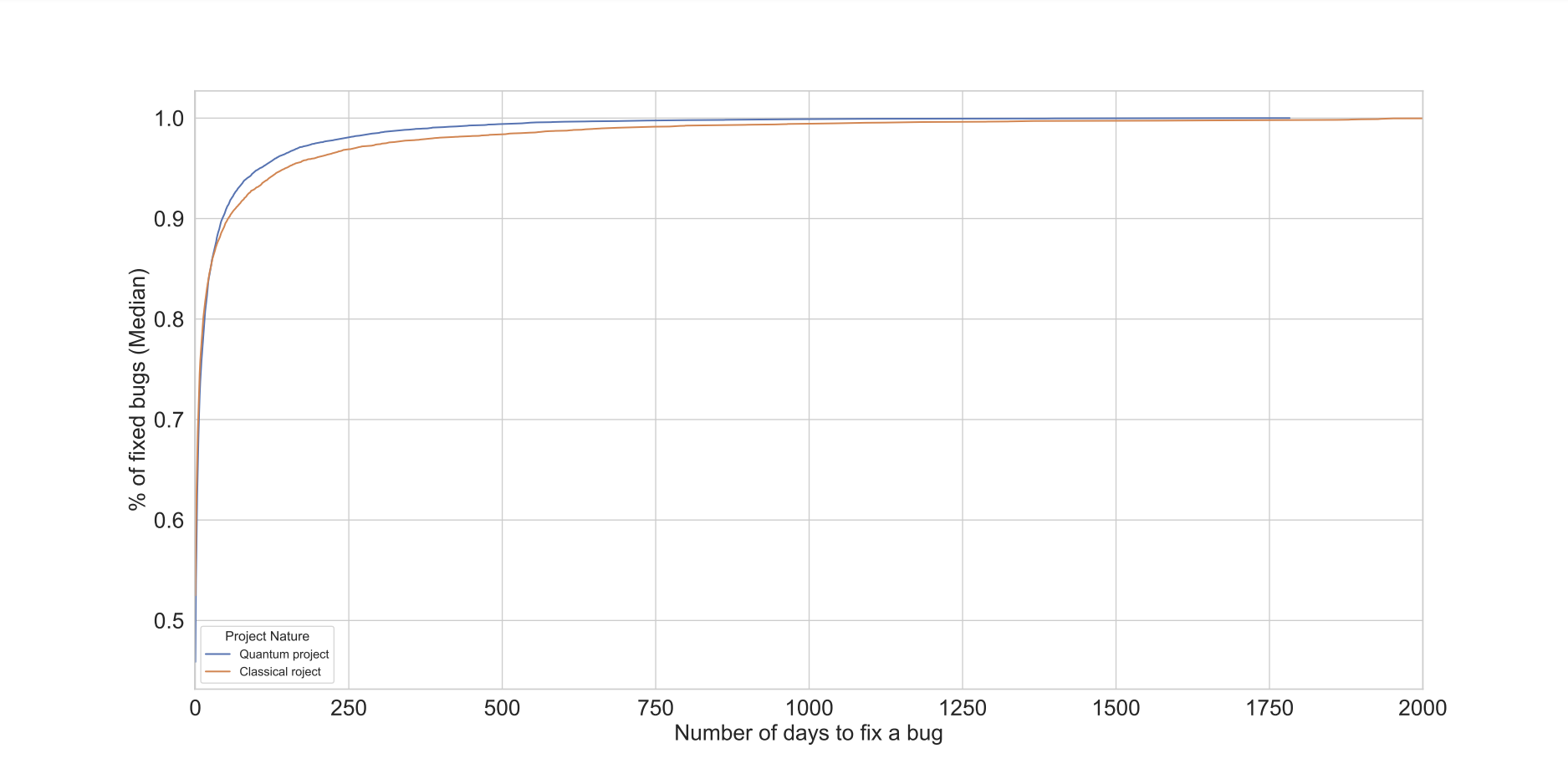}
  \captionof{figure}{Cumulative distribution function of duration to fix a bug through time (days).}
  \label{figure:CDF}
\end{minipage}%
\vspace{-3mm}
\end{figure}

\textbf{Developers of quantum software projects are actively addressing the bugs in their projects.} Figure~\ref{figure:CDF} show a comparison of the empirical distribution function of the duration to fix a bug in days. The figure shows that bugs in quantum software projects are fixed at a similar speed as in classical software projects, with 50\% of bugs fixed in less than 1 day and 90\% of bugs fixed within 48 days, indicating that developers of quantum software projects are actively maintaining their projects.

\textbf{\texttt{Quantum machine learning}, \texttt{Quantum based-simulation}, and \texttt{Experi-\\mental quantum computing} are more buggy than other categories}. Table \ref{table:categories} shows the median number of bugs across quantum software projects categories and the cost to fix them. We observe that  \texttt{quantum based-simulation} projects have the highest median fixed bugs with the value of 406.0, however, we also observe a low median bug fix ratio compared to other categories even though this category has a high median number of contributors as observed in Section \ref{sec:prestudy}. \texttt{Quantum machine learning, quantum algorithms and application, and Experimental quantum computing} categories have the highest bug fixing rate with respectively a median bug fix proportion (ratio) of  0.736, 0.657, and 0.628, meaning that those two categories are the least difficult for developers to fix. The highest number of bugs is for \texttt{quantum-based simulation} projects. There are almost two times more bugs in this category of quantum programs in comparison to the other projects, indicating that these projects need more support from the community for testing and debugging. In the \texttt{quantum-based simulation} category, programs are trying to simulate quantum mechanics and physics in a quantum computer, for this purpose developers have to implement mathematical equations, which may be complex for developers; inducing more bugs.

\textbf{As quantum programming is still low-level, bug fixing is costly in terms of code changes compared to classical programming.} Fixing a bug in a quantum program requires larger code changes than fixing bugs in classical programs. Table~\ref{table:categories} shows that quantum programs have a median code line change of 19, while classical programs have a median code line change of 14.00. We attribute this situation to the low-level programming language used in quantum programming. Quantum software engineers write their code at the gate level which is similar to the logic gate for classical software. This produces more coupling between the elements in the code which can lead to more changes. Since quantum programming is probabilistic in nature, classic assertions and testing techniques can't be used; resulting in a lack of support for testing and debugging.

\textbf{Bugs in \texttt{Quantum machine learning}, \texttt{Quantum programming framework}, \texttt{Quantum-based simulation} are the most challenging to fix}. Duration is the time difference between when an issue report was created and closed. In Table~\ref{table:categories}, we observe that the projects from \texttt{Quantum machine learning, Quantum programming frameworks, and Quantum-based simulation} categories appear to take the longest time to fix with the respective median number of hours to fix the bug of 66.05, 34,90, and 33,94. Also, the category \texttt{Quantum based-simulation} has the highest median number of lines of code changed in a commit with 63.00 median code line changed. Moreover, \texttt{Quantum machine learning} and \texttt{Quantum programming frameworks} show a high median number of lines of code changed to fix a bug. 

\begin{tcolorbox}
The quantum software projects are more buggy than classical software projects. While developers of quantum software projects are actively addressing their bugs,
fixing quantum software bugs is more costly than fixing classical software bugs.
Our results indicate the need for efforts to help developers identify quantum bugs in the early development phase (e.g., through static analysis), to reduce bug reports and the cost of bug fixing. Projects in the \texttt{quantum machine learning}, \texttt{quantum programming framework}, and \texttt{quantum-based simulation} categories are the most buggy and have the most difficult bugs to fix.
\end{tcolorbox}

\vspace{-5pt}
\subsection{\textbf{What are the characteristics of quantum software bugs?}}
\setcounter{subsubsection}{0}
\subsubsection{\textbf{Motivation}}
Quantum programs' execution flow is different from that of classical programs. To understand the challenges in the quantum computing ecosystem,  
it is important to understand how bugs are distributed among the different components of the programs.

\subsubsection{\textbf{Approach}}
To identify the different types of bugs occurring in quantum programs and the principal components of the quantum program execution flow where the bugs occurred, 
we performed a manual analysis of a statistically representative sample of bugs detected in RQ1. We performed a stratified sampling over the quantum program categories, to ensure covering a wide range of bugs. 
Specifically, 1) we compute the sample size from the number of fixed bugs with a 95\% confidence level and 5\% interval which result in a sample size of 376. 2) We calculate the log number of issues per category. 3) We compute the project category weight: the normalized log number of bugs of the category divided by the log of the total number of fixed bugs. 4) Finally, we compute the sample size from each project category: the total sample size (376) multiplied by the corresponding weight.

After creating the sample, for each bug report, we examine the title, body, and comments to understand the bug reported by the user. We used the hybrid-card sorting approach to perform the manual analysis and assign labels (i.e., type of bug and quantum program component) to each sampled bug. To assign the bug type, we based our manual analysis on an existing taxonomy of the type of bugs reported on GitHub~\cite{CATOLINO2019165} and added new types when needed.
For each bug report, we assigned one label, in case a bug is associated with two or more labels, which we found only in a few cases, we choose the most relevant one. 

We performed the labeling through two rounds as follows:
\begin{enumerate}
    \item \textbf{First-round labeling.} One coder labels the bugs independently.
    \item \textbf{First-round discussion.} In order to have a consistent labeling strategy, the first three authors of the paper had a meeting to discuss the labeling results in the first round and reached an agreed-upon set of labels.
    \item \textbf{Revising first-round labels.} The coder updated the first round labeling results based on the meeting.
    \item \textbf{Second-round discussion.} The first three authors of the paper had another meeting to discuss the revised labels, validate the updated labels and verify the consistency of the labels.
    \item \textbf{Revising second-round labels.} Based on the second-round discussion, the coder revised the labels.
\end{enumerate}

\subsubsection{\textbf{Results}}
\begin{figure}[!t]
\vspace{1mm}
\centering
\centerline{\includegraphics[width=0.96\textwidth]{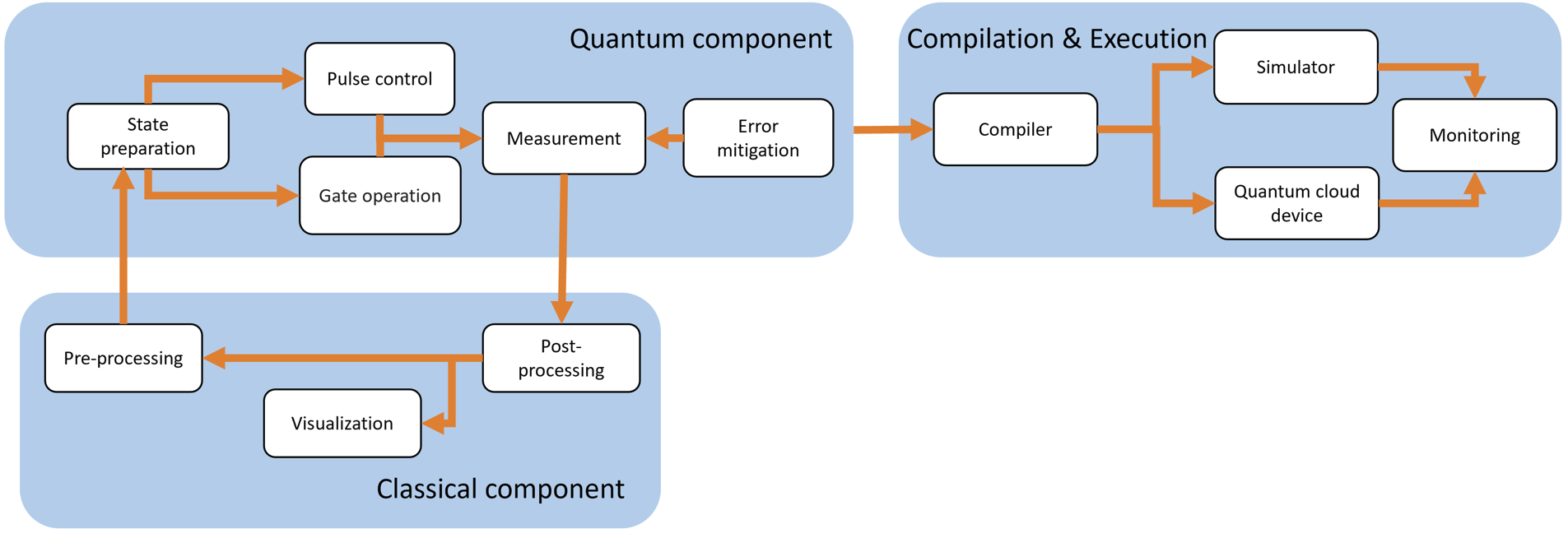}}
\centering
\caption{General components in a hybrid quantum program application}
\vspace{-2mm}
\label{figure:component}
\end{figure}

\begin{table*}[]
\tiny
\caption{Distribution of the quantum component bugs across the quantum software categories. (In the 376 analyzed bug reports, 43 reports were not identified as bugs.)}
\resizebox{.99\textwidth}{!}{
\begin{tabular}{@{}l|ccccccccccccc|c|@{}}
\cmidrule(l){2-15}
 &
  \multicolumn{1}{l}{VIS} &
  \multicolumn{1}{l}{SP} &
  \multicolumn{1}{l}{GO} &
  \multicolumn{1}{l}{Post-P} &
  \multicolumn{1}{l}{Pre-P} &
  \multicolumn{1}{l}{EM} &
  \multicolumn{1}{l}{CP} &
  \multicolumn{1}{l}{MS} &
  \multicolumn{1}{l}{PC} &
  \multicolumn{1}{l}{SM} &
  \multicolumn{1}{l}{MN} &
  \multicolumn{1}{l}{QCA} &
  \multicolumn{1}{l|}{O} &
  \multicolumn{1}{l|}{\# bugs} \\ \midrule
\multicolumn{1}{|l|}{\begin{tabular}[c]{@{}l@{}}Quantum programming\\ framework\end{tabular}}   & \checkmark  & \checkmark & \checkmark  & \checkmark  & \checkmark & \checkmark  & \checkmark & \checkmark  & \checkmark  &\checkmark  &  &   & \checkmark  & 58  \\
\multicolumn{1}{|l|}{\begin{tabular}[c]{@{}l@{}}Experimental quantum\\ computing\end{tabular}} & \checkmark  & \checkmark  & \checkmark  &   &  &   & \checkmark & \checkmark  & \checkmark  & \checkmark  &  & \checkmark  & \checkmark  & 47  \\
\multicolumn{1}{|l|}{Quantum   algorithms}             & \checkmark  & \checkmark  & \checkmark & \checkmark  & \checkmark & \checkmark  & \checkmark  & \checkmark  &   &   &  & \checkmark  & \checkmark  & 34  \\
\multicolumn{1}{|l|}{\begin{tabular}[c]{@{}l@{}}Quantum circuit \\ simulators\end{tabular}}     & \checkmark  &\checkmark  & \checkmark  &\checkmark  & \checkmark & \checkmark  & \checkmark  & \checkmark  & \checkmark  & \checkmark & \checkmark &   & \checkmark  & 43  \\
\multicolumn{1}{|l|}{Quantum   compilers}              &   &   & \checkmark  &   &  &   & \checkmark & \checkmark  & \checkmark  &   &  &   & \checkmark  & 37  \\
\multicolumn{1}{|l|}{\begin{tabular}[c]{@{}l@{}}Quantum utility\\ tools\end{tabular}}          & \checkmark  &    & \checkmark  & \checkmark  & \checkmark & \checkmark  & \checkmark  &   & \checkmark  & \checkmark &  & \checkmark & \checkmark  & 41  \\
\multicolumn{1}{|l|}{\begin{tabular}[c]{@{}l@{}}Quantum-based\\ simulation\end{tabular}}       & \checkmark  & \checkmark  & \checkmark  & \checkmark  & \checkmark &   & \checkmark  &   & \checkmark  &   & \checkmark & \checkmark  & \checkmark & 38  \\
\multicolumn{1}{|l|}{\begin{tabular}[c]{@{}l@{}}Quantum   Machine\\ learning\end{tabular}}       & \checkmark  & \checkmark  &\checkmark  & \checkmark  & \checkmark & \checkmark  & \checkmark  & \checkmark  &   &   &  & \checkmark  & \checkmark  & 35  \\ \midrule
\multicolumn{1}{|l|}{Total Bugs}                       & 15 & 40 & 56 & 17 & 9 & 15 & 68 & 19 & 17 & 10 & 2 & 25 & 40 & 333 \\ \bottomrule
\end{tabular}
}
\label{table:component}
\vspace{-2mm}
\end{table*}

\textbf{We identified bugs in 12 quantum computing components}.
Figure~\ref{figure:component} shows the 12 quantum computing computing components where bugs were identified which are mapped to a quantum-classical hybrid computing structure. Prior work~\cite{Sodhi2018QualityAO} identified five components : \texttt{pre-processing}, \texttt{post-pro-\\cessing}, \texttt{gate operation}, \texttt{state preparation}, and \texttt{measurement}. In Figure~\ref{figure:component}, we extended the component set to finally have 12 components. In Table \ref{table:bugs_type}, we show the distribution of the bug type across the quantum components. In the following, we discuss each quantum component in more details.

 \begin{itemize}
     \item \textbf{\texttt{Compiler (CP)}} translates the quantum program circuit into device-level language. From Table \ref{table:component}, we observe that compiler component issues are present in all the quantum ecosystem software categories with a total of 68 bugs. From table \ref{table:bugs_type}, we observe that the four most occurring bug types are program anomaly (23 bugs), test-code related (12 bugs), data type and structures (11 bugs), and configuration (11 bugs).

     \item \textbf{\texttt{Gate operation (GO)}} is a set of reversible operations that alter the state of the qubit and generate superposition. These reversible gates can be represented as a matrix. Gate operations are based on complex mathematical operations using imaginary numbers. Translating the maths into code is challenging for developers which increases the chances of errors. Table \ref{table:component} shows that \texttt{gate operations} component is the second most buggy with 56 bugs detected in our sample. Table \ref{table:bugs_type} illustrate that in this component, we have the two occurring bug types which are \texttt{program anomaly} and \texttt{data-type and data structure}. They respectively occur 27 and 7 times.
     
     \item \textbf{\texttt{Simulator (SM)}} simulates the execution of a circuit on a quantum computer. Since quantum computers are still not conveniently accessible by the public, simulators are widely used by quantum developers. We identify 10 bugs spread across the \texttt{quantum programming framework}, \texttt{experimental quantum computing},  \texttt{quantum circuit simulators}, and \texttt{quantum uti-\\lity tools}. We observe that the bugs are characterized mostly by \texttt{missing Error handling} and \texttt{performance bugs} with 2 bugs at each type as shown in table \ref{table:bugs_type}.
     
     \item \textbf{\texttt{State preparation (SP)}} is an operation to generate an arbitrary state of the qubit. As can be seen in Table \ref{table:component}, state preparation has the third-highest number of bugs with 40 bugs. The bugs appear in all the project categories except \texttt{quantum compilers} and \texttt{quantum utility tools} categories. During the state preparation phase, one tries to give the qubit an arbitrary state of 0 or 1. For this purpose, one tries to rotate or control the qubit into the desired state. As an example, one can use gate decomposition as a technique for state preparation. In Table \ref{table:bugs_type}, we observe that program anomaly is the most dominant bug type with 21 bugs, while data type and structure come second with only 5 bugs. However, we detected a rare bug of missing error handling type. Quantum states are mathematical entity that provide a probability distribution for each outcome of a state measurement. Since a qubit can have multiple states, mathematically they are represented as matrices. To initialize the state of a qubit, one has to manipulate that matrix. The number of state preparation bugs indicates the difficulty that developers are having with implementing the quantum algorithms. These bugs can be addressed with the help of data refinement and verification tools. Also, missing error handling bugs can be avoided by encouraging quantum developers to test their code.
     
     \item \textbf{\texttt{Measurement (MS)}} measures the final state of a qubit after superposition. We identified 19 bugs present in six of the quantum software categories as presented in Table \ref{table:component}. Six bug types were detected in our sample. From table \ref{table:bugs_type} \texttt{Program anomaly bugs} are the most occurring with 12 bugs. Also, we identified 3 \texttt{configuration bugs} and 2 \texttt{data type and structure bugs}. 
     \item \textbf{\texttt{Post-processing (Post-P)}} translates the quantum information (i.e., measurement probability) to classical computer bits. We detected 17 bugs related to \texttt{post-processing} distributed among all the quantum software categories except \texttt{experimental quantum computing} and \texttt{quantum compi-\\lers} as highlighted in the table \ref{table:component}. In table \ref{table:bugs_type} we observe 9 \texttt{program anomaly} bug and 3 \texttt{gui related} bugs. 

     \item \textbf{\texttt{Pre-processing (Pre-P)}} generates the state preparation circuit and initializes the register of the quantum computer. This component is executed on classical computer.  We detected 17 bugs related to \texttt{post-processing} distributed among all the quantum software categories except \texttt{experimen-\\tal quantum computing} and \texttt{quantum compilers} as highlighted in Table \ref{table:component}. In Table \ref{table:bugs_type}, we observe 9 \texttt{program anomaly} bugs and 3 \texttt{GUI related} bugs. 
     
     \item \textbf{\texttt{Pulse control (PC)}} generates and controls signals to create custom gates and calibrate the qubits. From Table \ref{table:component}, the pulse control component registered 17 bugs during the manual analysis. Pulse control bugs generally appear when a developer would like to define a custom gate and manually calibrate the qubits. For this purpose, developers need to generate a pulse (signal) and a scheduler to calibrate the qubit. From Table \ref{table:bugs_type}, it appears that the two most occurring bug types in pulse control components are \texttt{program anomaly} and \texttt{test-code related}. 
     
     \item \textbf{\texttt{Error mitigation (EM)}} corrects the measurement error when computing the qubit state probability. From Table \ref{table:component} we identified 15 error mitigation bugs distributed along 5 quantum software categories. Among the identified bugs we have detected 5 \texttt{program anomaly} bugs and 4 \texttt{test-code related bugs} as shown in Table \ref{table:bugs_type}.
     
     \item \textbf{\texttt{Quantum cloud access (QCA)}} is a client interface to connect to a quantum cloud service provider. Cloud service providers like Amazon, Microsoft, and IBM provide 
     access to quantum computers. We have detected 25 bugs related to the access to these quantum resources. \texttt{Configuration bugs} are the most occurring bug type with 8 bugs and \texttt{program anomaly bugs} are the second most frequent bug type in this category, with 5 bugs as shown in Table\ref{table:bugs_type}.
     
     \item \textbf{\texttt{Visualisation (VIS)}} creates plots and visualisations. Quantum developers use visualization to represent qubit states graphically and examine quantum state vectors and the transformation actions. Drawing the circuit is the last step when building a quantum program. We identified 15 bugs in the visualization component, that appear in all the quantum programs categories except the \texttt{quantum compilers}. Among the detected bugs, 8 belong to the \texttt{GUI related bugs} category as shown in Table \ref{table:bugs_type}. 
     
     \item \textbf{\texttt{Monitoring (MN)}} controls the good execution of the program and the performance of the execution environment. Only 2 bugs were assigned to the monitoring component, which appears in two quantum software categories (Quantum circuit simulation and Quantum-based simulation). 
     
     \item \textbf{\texttt{Other (O)}}: Every bug that occurred elsewhere other than the defined components. 
 \end{itemize}
 
The  result  of  our  qualitative analysis for identifying the bugs in the quantum program software component is presented in the Table \ref{table:component}. Among the 376 analyzed bug reports, 43 questions reported bugs were not real bugs.

We identify 147 bugs in 5 components that are directly related to quantum computing including \texttt{gate operation}, \texttt{state preparation}, \texttt{measurement},\\ \texttt{pulse control}, and \texttt{error mitigation}, ordered by their number of bugs. Moreover, the \texttt{simulator} shows a slightly high number of bugs (classical and quantum bugs), this can be explained by the complexity of the operation that has to be simulated since they are trying to simulate quantum physic phenomena using very complex linear algebra calculation.
Finally, the \texttt{compilers} appear to be the most buggy with 63 bugs identified. 

\begin{tcolorbox}
We identified a variety of 12 quantum computing components in which bugs were discovered, including both quantum computing-related bugs and classical-related bugs. We identified quantum-related bugs in five components (\texttt{gate operation}, \texttt{state preparation}, \texttt{measurement}, \texttt{pulse control}, and \texttt{error mitigation}.), among which \texttt{gate operation} is the most buggy component.
\end{tcolorbox}
\begin{table*}[]
\tiny
\caption{Distribution of bug types across the quantum component}
\resizebox{.99\textwidth}{!}{
\begin{tabular}{l|ccccccccccccc|c|}
\cline{2-15}
\textbf{} &
  \multicolumn{1}{l}{VIS} &
  \multicolumn{1}{l}{SP} &
  \multicolumn{1}{l}{GO} &
  \multicolumn{1}{l}{Post-P} &
  \multicolumn{1}{l}{Pre-P} &
  \multicolumn{1}{l}{EM} &
  \multicolumn{1}{l}{CP} &
  \multicolumn{1}{l}{MS} &
  \multicolumn{1}{l}{PC} &
  \multicolumn{1}{l}{SM} &
  \multicolumn{1}{l}{MN} &
  \multicolumn{1}{l}{QCA} &
  \multicolumn{1}{l|}{O} &
  \multicolumn{1}{l|}{\# bugs} \\ \hline
\multicolumn{1}{|l|}{Configuration bugs}              & 3  & 3  & 4  & -  & -  & 2 & 11 & 3  & 1  & 2  & -  & 8  & 10 & 47  \\
\multicolumn{1}{|l|}{Data types and structures bugs} & 3  & 5  & 10  & 1  & 3  & - & 12 & 2  & 2  & 1  & -  & 2  & 5  & 46  \\
\multicolumn{1}{|l|}{Missing Error handling}          & -  & 1  & -  & -  & -  & 1 & 4  & -  & 3  & 2  & -  & 2  & -  & 12   \\
\multicolumn{1}{|l|}{Performance bugs}                & -  & -  & 3  & 1  & 1  & 3 & 3  & 1  & 1  & 2  & -  & 1  & 3  & 19  \\
\multicolumn{1}{|l|}{Permission/deprecation bugs}     & -  & -  & 2  & 1  & -  & - & 1  & -  & -  & -  & -  & 1  & 2  & 7   \\
\multicolumn{1}{|l|}{Program anomaly bugs}            & 2  & 20 & 30 & 9  & 4  & 5 & 26 & 12 & 6  & 1  & -  & 5  & 2  & 123 \\
\multicolumn{1}{|l|}{Test code-related bugs}          & -  & 2  & 4  & -  & -  & 4 & 12 & 1  & 4  & -  & 1  & -  & 5  & 33  \\
\multicolumn{1}{|l|}{DataBase related bugs}          & -  & -  & -  & -  & -  & - & -  & -  & -  & -  & -  & -  & 1  & 1   \\
\multicolumn{1}{|l|}{Documentation}                    & -  & 1  & 3  & -  & -  & - & -  & -  & -  & 1  & -  & 1  & 11 & 17  \\
\multicolumn{1}{|l|}{Gui related bugs}               & 8  & 3  & 1  & 3  & -  & - & -  & -  & -  & -  & -  & -  & -  & 15  \\
\multicolumn{1}{|l|}{Misuse}                           & -  & 2  & 1  & -  & -  & - & -  & 1  & -  & -  & -  & 1  & -  & 2   \\
\multicolumn{1}{|l|}{Network bugs}                   & -  & -  & -  & -  & -  & - & -  & -  & -  & -  & -  & 1  & -  & 1   \\
\multicolumn{1}{|l|}{Monitoring}                       & 1  & -  & -  & -  & -  & - & -  & -  & -  & -  & -  & -  & -  & 1   \\ \hline
\end{tabular}
}
\label{table:bugs_type}
\end{table*}

We identified \textbf{13 bug types in the quantum software ecosystem}. Table~\ref{table:bugs_type} shows the distribution of these bugs across quantum components. In the following, we discuss each type of bugs in more details. In our replication package~\cite{ReplicationPackage}, we provided more details and examples of different types of bugs in different quantum computing components.

\textbf{\texttt{Program anomaly bugs}} are introduced when enhancing existing code or caused by prior bad implementations. The manifestation of these bugs can be 
a bad return value or unexpected crashes due to logical issues in the code. For example in the \texttt{Pre-processing} component, to execute quantum programs efficiently, data is embedded into quantum bits. Specifically, the classical data is mapped into n-qubit quantum states by transforming data into a new space where it is linear. For example, AmplitudeEmbedding is a mapping technique in \texttt{PennyLaneAI} in which a bug\footnote{\url{https://github.com/PennyLaneAI/pennylane/issues/365}} was reported and it took 6 days to fix. In the code from Listing \ref{lst:pre_pa}(a), the normalization in the embedding function is breaking the differentiability for any data that is encoded as a \texttt{TensorFlow} or \texttt{Pytorch} tensor which triggers the error shown in Listing \ref{lst:pre_pa}(b). Developers have tried pre-implemented normalization techniques in \texttt{NumPy} and \texttt{TensorFlow} but the bug persisted. The only solution left was to hand-code the normalization as illustrated in the bug fix code from Listing \ref{lst:pre_pa}(c). Providing the quantum computing community with libraries dedicated to linear algebra computation in quantum will support the progress of quantum programming.
\begin{figure}[!ht]
\captionsetup{type=lstlisting}

\begin{sublstlisting}{\linewidth}
\begin{lstlisting}[language=Python]
import numpy as np
import pennylane as qml
from pennylane.templates.embeddings import AmplitudeEmbedding

dev = qml.device('default.qubit', wires=3)
@qml.qnode(dev)
def circuit(data):
#Amplitude embedding normalization function not behaving as expected (Bug trigger)
    AmplitudeEmbedding(data, wires=[0, 1, 2])
    return qml.expval.PauliZ(0)

data = np.ones(shape=(8,)) / np.sqrt(8)
circuit(data)
\end{lstlisting}
\caption{Trigger of the bug "PennyLaneAI (issue id 365)"}
\end{sublstlisting}

\begin{sublstlisting}{\linewidth}
\begin{lstlisting}[language=Python]
TypeError: unsupported operand type(s) for +: 'Variable' and 'Variable'
\end{lstlisting}
\caption{Error message}
\end{sublstlisting}
\vspace{-3mm}

\begin{sublstlisting}{\linewidth}
\begin{lstlisting}[language=Python]
#Repository: PennyLaneAI/pennylane
#Fix file: pennylane/templates/embeddings.py, method: AngleEmbedding, line: 118
-if normalize and np.linalg.norm(features, 2) != 1:
-    features = features * (1/np.linalg.norm(features, 2))
+norm = 0
+for f in features:
+if type(f) is Variable:
    +norm += np.conj(f.val)*f.val
+else:
    +norm += np.conj(f)*f        
\end{lstlisting}
\caption{Proposed bug fix}
\end{sublstlisting}
\vspace{-3mm}

\caption{The execution of the code snippet (shown in (a)) triggers a message error (shown in (b)) because Amplitude embedding normalization in NumPy is too strict and does not allow small tolerance. The bug fix was to redefine the normalization function (as shown in (c))}
\label{lst:pre_pa}
\vspace{-3mm}
\end{figure} 
    
\textbf{\texttt{Test-code related bugs}} are happening in the test code. Problem reported due to adding or updating test cases and intermittently executed tests. For example, in the pulse control component, \texttt{Test code-related bug} is the second most occurring bug type. In \texttt{Qiskit} bug report number 2527\footnote{\url{https://github.com/Qiskit/qiskit-terra/issues/2527}}, a unitest is failing because of a wrong parameter type. The expected input variable type in this test should be int, float, or complex. Therefore to fix the bug, a developer changed the parameter type from String to Integer as shown in Listing \ref{lst:pc_tst_fix}.
\begin{figure}[!ht]  
    \begin{lstlisting}[language=Python, caption=The execution of the test building parameterized schedule fails because of the wrong parameter type (as shown in code snippet line 1 and 5). The bug fix was to change the parameter type from string to integer ``qiskit (issue id 2527)'',label=lst:pc_tst_fix]
    #Repository: Qiskit/qiskit-terra
    #Fix File: qiskit-terra/test/python/pulse/test_cmd_def.py
    #Class: TestCmdDef, method:test_parameterized_schedule, line: 86
    #Bug triggered in line 4 and 8
    -sched = cmd_def.get('pv_test', 0, '0', P2=-1)
    +sched = cmd_def.get('pv_test', 0, 0, P2=-1)
    self.assertEqual(sched.instructions[0][-1].command.value, -1)
    with self.assertRaises(PulseError):
        -cmd_def.get('pv_test', 0, '0', P1=-1)
        +cmd_def.get('pv_test', 0, 0, P1=-1)
    \end{lstlisting}
\vspace{-3mm}
\end{figure}

\textbf{\texttt{Data type and structure bugs}} are related to  data type problems such as undefined or mismatch type and data structure bugs like bad shape bugs or the use of wrong data structures. For example, in \texttt{qibo}, we identified a bug that states : ``\texttt{\textbf{Probabilities do not sum to 1}}\footnote{\url{https://github.com/qiboteam/qibo/issues/517}}" when running a circuit with more than 25 qubits. Listing \ref{lst:ms_dt}(a) presents the buggy code along with the error message and the proposed bug fix. As can be seen, the problem has occurred because the \texttt{sum()} function does not return a normalized set of probabilities distribution. To fix the problem, a developer proposed to cast the tensor into type \texttt{tf.complex128}. This bug was caused by a data type and structure bug in the \texttt{measurement} component. Quantum software programs algorithms are based on probability and linear algebra which makes the implementation of quantum algorithms complex for developers. Therefore, providing libraries with array and data manipulation for quantum and data validation tools can help the quantum software engineering community.
\begin{figure}[!ht]
\captionsetup{type=lstlisting}
\begin{sublstlisting}{\linewidth}
\begin{lstlisting}[language=Python]
NQUBITS = 26
c = Circuit(NQUBITS)
for i in range(NQUBITS):
    c.add(gates.H(i))
output = c.add(gates.M(TARGET, collapse=True))
for i in range(NQUBITS):
    c.add(gates.H(i))
# Running the circuit with more than 25 qubit return sum of qubit states probabilities =! 1 (Bug triggered)
result = c()
\end{lstlisting}
\caption{Trigger of the bug "qibo (issue id 517)"}
\label{lst:ms_dt_a}
\end{sublstlisting}
\begin{sublstlisting}{\linewidth}
\begin{lstlisting}[language=Python]
File "mtrand.pyx", line 933, in numpy.random.mtrand.RandomState.choice
ValueError: probabilities do not sum to 1
\end{lstlisting}
\caption{Error message}
\end{sublstlisting}
\begin{sublstlisting}{\linewidth}
\begin{lstlisting}[language=Python]
#Repository:qiboteam/qibo
#Fix file: qibo/src/qibo/core/states.py
#Class:VectorState, method: probabilities(), line: 95
def probabilities(self, qubits=None, measurement_gate=None):
    unmeasured_qubits = tuple(i for i in range(self.nqubits) if i not in qubits)
-    state = K.reshape(K.square(K.abs(K.cast(self.tensor))), self.nqubits * (2,))
+    state = K.reshape(K.square(K.abs(K.cast(self.tensor, dtype="complex128"))), self.nqubits * (2,))
    return K.sum(state, axis=unmeasured_qubits) 
\end{lstlisting}
\caption{Bug fix}
\end{sublstlisting}
\vspace{-2mm}
\caption{The execution of the code snippet (shown in (a)) triggers the message error (shown in (b)) because the sum of final state probabilities is different from 1. The bug fix was to cast the array value to complex type (as shown in (c))}
\label{lst:ms_dt}
\vspace{-5mm}
\end{figure}
    
\textbf{\texttt{Missing error handling}} occur when exceptions are not handled by the program. 
Improper error handling can lead to serious consequences for any system, and the quantum software systems are not an exception. For example, in \texttt{qutip}, we detect the absence of error handling in the simulator. An issue that led to the bug report id 396 \footnote{\url{https://github.com/m-labs/artiq/issues/396}}. The application crashed during the creation of a device object if the methods or attributes of this object contained an error.  
This issue was fixed by adding an exception block as shown in Listing \ref{lst_sm_er}. This bug fix occurred quickly (1 day) and not much code change was required. Improper error handling can be costly and lead to data leaks and many other exploits in the code. Therefore, developers must be careful regarding when, where, and how to correctly handle errors in the code. Code analysis tools for error handling and logging recommendation can support the quantum program development and help avoid \texttt{missing error handling} bugs. 
       
\begin{figure}[!ht]
\captionsetup{type=lstlisting}  
\begin{lstlisting}[language=Python, caption=Calling the device description causes the program to crash with inappropriate error message hard to understand.The bug fix was adding exception block in the get method with descriptive error message "artiq (issue id 396)".,label=lst_sm_er]
#Repository: m-labs/artiq
#Fix file: artiq/master/worker_db.py 
#Class: DeviceManager, method: get, line: 144
-dev = _create_device(self.get_desc(name), self)
+try:
+    desc = self.get_desc(name)
+except Exception as e:
+    raise DeviceError("Failed to get description of device '{}'".format(name)) from e
+try:
+    dev = _create_device(desc, self)
+except Exception as e:
+    raise DeviceError("Failed to create device '{}'".format(name)) from e
\end{lstlisting}
\vspace{-5mm}
\end{figure}
    
\textbf{\texttt{Configuration bugs}} are related to configuration files building. The bug can be caused by the external library that must be updated or incorrect files paths or directories. For example, in the \texttt{Quantum cloud access} component, The bug in the code from Listing \ref{lst:qca_cf} happened in \texttt{DWave cloud access service} when trying to obtain the list of solvers. However, the list of solvers is filtered by the client type. To locate and fix the bug, developers spent 
14 days \footnote{\url{https://github.com/dwavesystems/dwave-cloud-client/issues/457}}. Their proposed fix is 
shown in Listing \ref{lst:qca_cf}.
\begin{figure}[!ht]    
\begin{lstlisting}[language=Python, caption=Dwave solver is sensitive to the client type and does not return all the solvers with option -all because developers missed the client type argument in the solvers function. The bug fix was passing the client type as argument "dwave-cloud-client( issue id 457) , label=lst:qca_cf]
#Repository:dwavesystems/dwave-cloud-client
#Fix file: dwave/cloud/cli.py, method: solvers, line:346
+@click.option('--client', 'client_type', default=None,
              type=click.Choice(['base', 'qpu', 'sw', 'hybrid'], case_sensitive=False),
              help='Client type used (default: from config)')
-def solvers(config_file, profile, solver_def, list_solvers, list_all):
+def solvers(config_file, profile, client_type, solver_def, list_solvers, list_all):
    if list_all:
+        client_type = 'base'
    with Client.from_config(
-            config_file=config_file, profile=profile, solver=solver_def) as client:
+            config_file=config_file, profile=profile,
+            client=client_type, solver=solver_def) as client:
\end{lstlisting}
\vspace{-5mm}
\end{figure}

\textbf{\texttt{GUI related bugs}} are related to graphical elements 
such as layout, text box, and button layout. Quantum circuits are the main focus when building a quantum program. It can be challenging to debug and--or validate the structure of a quantum circuit by only reading the code. Libraries like \texttt{Cirq} and \texttt{Qiskit} offer visualization features to draw the architecture of circuits. The drawing is meant to depict the physical arrangement of gates, wires, and components. The visualization functions in \texttt{Qiskit} are basic support modules and can come with bugs. For example, the bug report\footnote{\url{https://github.com/Qiskit/qiskit-terra/issues/3107}} in \texttt{Qiskit} is stating misaligned barriers in the circuit, which change the interpretation of the circuit. In the code presented in Listing \ref{lst:pst_gui}, we show a bug in Qiskit that occurred when a user wanted to draw a circuit, alongside with its proposed fix. 
\begin{figure}[!ht]
\captionsetup{type=lstlisting}

\vspace{-3mm}
\begin{sublstlisting}{\linewidth}
\begin{lstlisting}[language=Python]
qc = qk.QuantumCircuit(2)
#The circuit draw in latex places barrier in the wrong place
qc.draw(output='latex')
\end{lstlisting}
\caption{Trigger of the bug qiskit (issue id 3107)}
\end{sublstlisting}
\vspace{-1mm}

\begin{sublstlisting}{\linewidth}
\begin{lstlisting}[language=Python]
#Repository: Qiskit/qiskit-terra
#Fix file:  qiskit/visualization/latex.py
#Class: QCircuitImage, method: _build_latex_array, line:772
-self._latex[start][column] = "\\qw \\barrier{" + str(span) + "}"
+self._latex[start][column - 1] += " \\barrier[0em]{" + str(span) + "}"
+self._latex[start][column] = "\\qw"
\end{lstlisting}
\caption{Trigger of the bug}
\end{sublstlisting}
\caption{The execution of the code snippet (shown in (a)) plots a circuit with a barrier not aligned correctly. The bug fix was to adjust the barrier column index (shown in (b))}
\label{lst:pst_gui}
\end{figure}
Libraries dedicated to quantum circuit visualization such us \texttt{Matplotlib} and \texttt{Seaborn} in python help developers to analyze and debug the circuits.

\textbf{\texttt{Performance bugs}} are related to the stability, speed, or response time of software resources. This category covers memory overuse, endless loops, and energy consumption. For example, in the \texttt{Qiskit} simulator, we found a bug that caused a huge performance regression in the simulation due to a large increase in serialization overhead when loading noise models from Python into C++. Even though the bug fix consisted only in one line of code, developers spent 4 days to locate and fix this bug. Quantum developers make a high use of quantum simulators, therefore, it is important to ensure that these simulators maintain a good performance. Creating a performance benchmark for quantum simulators can help the community to select the most appropriate simulators and encourage providers to improve the performance of their simulators.
\begin{figure}[!ht]
\captionsetup{type=lstlisting}
\begin{lstlisting}[language=Python]
#qiskit/qiskit-aer
# Fix file:  qiskit/providers/aer/noise/errors/quantum_error.py, method: to_dict, line: 462
-instructions = [exp.to_dict()['instructions'] for exp in qobj.experiments]
+instructions: [[op[0].assemble().to_dict() for op in circ.data]
                             for circ in self._circs]
\end{lstlisting}
\caption{Calling qiskit.assemble causes a huge performance regression in noisy simulator "qiskit-aer (issue id 1398)"}
\label{lst:sm_prf}
\end{figure}
    
\textbf{\texttt{Permission/deprecation bugs}}. This category covers two type of bugs: (1) bugs related to the removal or modification of deprecated calls or APIs, and (2) bugs related to missing or incorrect API permission.
    
\textbf{\texttt{Database related bugs}} are related to the connection between the database and the main application.
    
\textbf{\texttt{Documentation}} issues are related to documentation typos, not up to date documentations and misleading function in documentation.

\textbf{\texttt{Monitoring bugs}} are related to bad logging practices such as wrong logging levels, too much logging, or missing log statement. Only 2 bugs were assigned to the monitoring component. For example, the bug report id 100\footnote{\url{https://github.com/sfc-aqua/quisp/issues/100}} in \texttt{quisp} describes a problem of overlogging and misplaced log statements. The overloaded text has been making 
debugging difficult. Even though, logging is a known practice in software engineering, its effectiveness requires adequate logging statements at appropriate locations in the code. 
Tools for logging and log level recommendations can help support quantum developers and improve the quality of their applications.

\textbf{A \texttt{Misuse}} is a wrong usage of a function that leads to a bug in the code. Misuses appeared in \textbf{gate operation}, \textbf{state preparation} components. For example, in the Listing \ref{lst:sp_pa}, we present a \textbf{Misuse} bug that occurred during the state preparation of the qubit. This bug occurred because the developer wrote the program in the wrong order. The purpose of the circuit presented in this example is to measure the state of qubit 0 and map the state into a classical bit. However, the program of this circuit raised a \texttt{CircuitError} as shown in Listing \ref{lst:sp_pa}(b). This 
error is due to line \texttt{QuantumCircuit()} which is not doing anything since it is immediately over-ridden by \texttt{statepreparation()}. Moreover, the measurement is not mapped to the register (classical bit) because of the misuse of the \texttt{measure()} function. A developer proposed the following fix to correct the issue (Listing \ref{lst:sp_pa}(c)).
\begin{figure}[!ht]
\captionsetup{type=lstlisting}
\begin{sublstlisting}{\linewidth}
\begin{lstlisting}[language=Python]
q = QuantumRegister(4)
c = ClassicalRegister(1)
circuit = QuantumCircuit(q,c)
ang = feature_train[3]
# Applying state preparation after quantum Circuit initialization triggers the bug
circuit = statepreparation(ang, circuit, [0,1,2,3])
circuit.measure(0, c)
\end{lstlisting}
\caption{Trigger of the bug "qiskit (issue Id 5837)"}
\end{sublstlisting}

\begin{sublstlisting}{\linewidth}
\begin{lstlisting}[language=Python]
CircuitError: 'register not in this circuit'
\end{lstlisting}
\caption{Code to reproduce the bug}
\end{sublstlisting}

\begin{sublstlisting}{\linewidth}
\begin{lstlisting}[language=Python, ]
#Fix file: In the code snipped (shown in (a)), line 3 and 7
-circuit = QuantumCircuit(q,c)
-circuit.measure(0, c)
+circuit.measure(0, 0)
\end{lstlisting}
\caption{Code to reproduce the bug}
\end{sublstlisting}

\caption{The execution of the code snipped (shown in (a)) triggers the message error (shown in (b)) because the quantum circuit (line 3 in code snipped (a)) is immediately overridden by the state preparation (line 6 in a code snipped (a)). The bug fix was to map the 0th measured qubit into the 0th classical bit and remove the circuit initialization (as shown in (c).}
\label{lst:sp_pa}
\end{figure}

\textbf{\texttt{Network bugs}} are related to connection or server issues. Network bug type is the rarest bug  Where we identify one bug in the quantum cloud access component from table \ref{table:bugs_type}. 

We identified 5 dominant bug type while doing the manual analysis: \texttt{program anomaly}, \texttt{data type and structure}, \texttt{configuration issues}, \texttt{test-code\\ related} and \texttt{enhancement and feature request}. \texttt{Program anomaly} is the most dominant bug type with 123 bugs. While inspecting the bug reports and their commit message, we notice that most program anomaly bugs come from bad implementation in the quantum algorithm or mathematical formulation of the problem. In fact, gate operation and state preparation components are based on complex mathematics, and this may be challenging for developers. \texttt{Configuration issues} is the second most frequent bug type with 45 bugs. Most of the quantum concepts are based on linear algebra and use complex data type; this makes the implementation challenging for developers. Data type and structure is the third most frequent category with 40 bugs. Again, this may be caused by the complexity of quantum operations.    

\begin{tcolorbox}
We identified 13 different types of quantum bugs. The most frequent types of bugs occurring in quantum components are: program anomaly bugs with 123 bugs, configuration bugs with 47 bugs, and data type and structure bugs with 46 occurrences. Program anomaly bugs is the most spread bug type (it occurs in all components). Data type and structure bugs are mostly located in the components \texttt{State preparation} (10 occurrences) and \texttt{Compiler} (12 occurrences).
\end{tcolorbox}

\noindent \textbf{Discussion on quantum bug categories.}
We have seen in the analysis that most of the bugs in the quantum software ecosystem are program anomalies, configuration bugs, and data type and structure bugs. These bugs can have a serious impact, causing the program to crash and leading to poor software quality. We observe that quantum programmers have limited or no access to data, matrix, and array manipulation libraries. It is often difficult for them to implement quantum algorithms, linear algebra routines, error mitigation models, and unitary transformation with existing libraries such as \texttt{Numpy}. This finding suggests that developing \texttt{\textbf{data manipulation (e.g., array manipulation) libraries}} dedicated to quantum programming along with a collection of \texttt{\textbf{mathematical algorithms for quantum computing and convenience fun-\\ctions}} can help prevent some of the observed 
program anomalies and data type and structure bugs. Our analysis also revealed the difficulty of designing quantum circuits. 
Automatic pattern recommendation tools could be developed to support this task. 
This can be done for example by mining a large set of expert-written quantum code to extract common code patterns that can be recommended to developers designing quantum circuits. 

Because quantum programs are difficult to debug and quantum components are strongly coupled, \textbf{\texttt{circuit visualization and analysis}} approaches are needed to help developers examine the logical structure of their circuits and fix bugs early on. 

\begin{tcolorbox}
Researchers and tool builders should consider contributing specialized \textbf{\texttt{data manipulation (e.g., array manipulation)}} libraries and libraries providing \textbf{\texttt{mathematical algorithms for quantum computing and convenience functions}} to support quantum software development, to help reduce the occurrence of program anomaly bugs and data type and structure bugs. \textbf{\texttt{Circuit visualization and analysis}} techniques and tools are also needed to help developers debug and fix bugs in quantum circuits.

\end{tcolorbox}

\section{Related Work} \label{sec:relatedwork}

We discuss prior work related to our study; organizing them in three categories: quantum software engineering, topic analysis of technical Q\&A data, and topic analysis of issue reports.

\subsection{Quantum Software Engineering}
Quantum software engineering (QSE) is still in its infancy. 
As the result of the first International Workshop on Quantum Software Engineering \& Programming (QANSWER), researchers and practitioners proposed the ``Talavera Manifesto" for quantum software engineering and programming, which defines a set of principles about QSE (e.g.,``QSE is agnostic regarding quantum programming languages and technologies'')~\cite{piattini2020talavera}.  
Zhao~\cite{zhao2020quantum} performed a comprehensive survey of the existing technology in various phases of the quantum software life cycle, including requirement analysis, design, implementation, testing, and maintenance. 
Prior work~\cite{moguel2020roadmap, piattini2020quantum, barbosa2020software, piattini2021toward} also discussed challenges and potential directions in QSE research, such as modeling~\cite{barbosa2020software} and quantum software processes \& methodologies~\cite{moguel2020roadmap}, and the design of quantum hybrid systems~\cite{piattini2021toward}. 
In addition, prior work conducted extensive exploration along the lines of quantum software programming~\cite{garhwal2019quantum} and quantum software development environments~\cite{larose2019overview}. The survey~\cite{zhao2020quantum} provides a comprehensive overview of the work along these lines.
Different from prior work, this work examines the characteristics of bugs occurring in the quantum software ecosystem. 

\subsection{Quantum Bugs Characteristics}
In their 2021 position paper, Campos and Souto~\cite{Campos2021QBugsAC} argued for the creation of a benchmark dataset of quantum bugs. In the same year, Zhao et al.~\cite{abs-2103-09069} provided a data set of 36 bugs identified in the quantum computing framework Qiskit. In 2022, Matteo and Michael~\cite{abs-2110-14560} examined 283 bugs from 18 open-source quantum computing platforms to identify bug patterns. In this paper, we study a larger set of bugs from a larger number of quantum projects. We perform quantitative and qualitative analysis on the characteristics and types of the quantum bugs occurring in different quantum components. 

\subsection{Quantum Programs Testing}
Quantum programs are more difficult to test and debug than a classical program because of the impossibility to copy the quantum information in the qubits~\cite{Wootters1982ASQ}, and the probabilistic nature of the measurement. To face these challenges, Huang and Martonosi~\cite{10.1145/3307650.3322213} introduced statistical assertions that can be used to validate patterns and detect bugs in quantum programs. Li et al.~\cite{10.1145/3428218} proposed \texttt{Proq}, a project-based runtime analysis tool for testing and debugging quantum programs. The evaluation of the tool shows that it can effectively help locate bugs in quantum programs. Yu and Palsberg~\cite{10.1145/3453483.3454061} proposed an abstract interpretation of quantum programs and use it to automatically verify assertions in polynomial time. Similar to our work, these previous works on quantum program testing contribute to improving our understanding of the nature of quantum bugs. 

\vspace{-4pt}
\section{Threats to Validity} \label{sec:threats}
We now discuss threats to the validity of our study.\\

\noindent \textbf{External validity.}
In this work, we analyze the bugs of 125 quantum software projects on GitHub. Our studied projects may not represent the characteristics of other quantum software projects that are not public on GitHub. In addition, our studied projects may not represent all quantum software projects on GitHub. However, we followed a systematic approach to search for quantum software projects and focus on projects with relatively rich development activities.

\noindent \textbf{Internal validity.}
In RQ1, we analyze the bug fixing efforts using the bug fix duration and the code changes in the bug fixes as proxies. However, the duration and code changes may not accurately capture the effort of developers in fixing these bugs. Future works that accurately monitor developers' development activities in fixing quantum software bugs can improve our analysis. 

\noindent\textbf{Construct validity.}
In our preliminary study and RQ2, we manually analyze the categories of the quantum software projects and the characteristics of the quantum software bugs. 
Our results may be subjective and depend on the judgment of the researchers who conducted the manual analysis. To mitigate this threat, two authors of the paper collectively conducted a manual analysis and reached a substantial agreement, indicating the reliability of the analysis results. To resolve disagreements, the third author joined them and each case was discussed until reaching a consensus. 

\noindent\textbf{Conclusion validity.} This threat concerns 
the extent to which the analyzed issues can be considered exhaustive enough. We have followed a systematic approach to identify the studied quantum projects. To identify the bug types and the components of the quantum program execution flow in which they occurred, we have manually analyzed a statistical representative sample of the detected bugs. Although it is possible that we may have missed some types of bugs, we have mitigated this threat by following a stratified sampling strategy that allowed us to cover projects from all quantum computing categories. 

\noindent\textbf{Reliability validity.} This threat concerns the possibility to replicate this study. We have attempted to provide all the necessary details needed to replicate our study. We share our full replication package in \cite{ReplicationPackage}.

\section{Conclusions} \label{sec:conclusions}

This work performs an empirical study on 125 open-source quantum software projects hosted on GitHub. These quantum software projects cover a variety of categories, such as quantum programming frameworks, quantum circuit simulators, or quantum algorithms. An analysis of the development activity of these selected projects show a level of development activities similar
to that of classical projects hosted on GitHub. We compared the distribution of bugs in quantum software projects and classical software projects, as well as developers’ efforts in addressing these bugs and observed that quantum software projects are more buggy than comparable classical software projects.
Besides, quantum software project bugs are more costly to fix (in terms of the code changed) than classical software project bugs. We qualitatively studied a statistically representative sample of quantum software bugs to understand their characteristics. We identified a total of 13 different types of bugs occurring in 12 quantum components. The three most occurring types of bugs are Program anomaly bugs, Configuration bugs, and Data type and structure bugs. These bugs are often caused by the wrong logical organization of the quantum circuit, state preparation, gate operation, measurement, and state probability expectation computation. Our study also highlighted the need for specialized data manipulation (e.g., array manipulation) libraries, libraries providing mathematical algorithms
for quantum computing and convenience functions, as well as circuit
visualization and analysis techniques and tools to support quantum software development.


\bibliographystyle{spbasic}
\bibliography{quantumcomp}

\begin{thebibliography}{44}
\providecommand{\natexlab}[1]{#1}
\providecommand{\url}[1]{{#1}}
\providecommand{\urlprefix}{URL }
\expandafter\ifx\csname urlstyle\endcsname\relax
  \providecommand{\doi}[1]{DOI~\discretionary{}{}{}#1}\else
  \providecommand{\doi}{DOI~\discretionary{}{}{}\begingroup
  \urlstyle{rm}\Url}\fi
\providecommand{\eprint}[2][]{\url{#2}}

\bibitem[{Aleksandrowicz et~al.(2019)Aleksandrowicz, Alexander, Barkoutsos,
  Bello, and al.}]{Qiskit}
Aleksandrowicz G, Alexander T, Barkoutsos P, Bello L, al (2019) {Qiskit: An
  Open-source Framework for Quantum Computing}.
  https://doi.org/10.5281/zenodo.2562111, {Accessed 2022-04-18}

\bibitem[{Aron(2019)}]{IBM_one}
Aron J (2019) Ibm unveils its first commercial quantum computer.
  \url{https://www.newscientist.com/article/2189909-ibm-unveils-its-first-commercial-quantum-computer/},
  accessed: 2022-01-16

\bibitem[{Barbosa(2020)}]{barbosa2020software}
Barbosa LS (2020) Software engineering for 'quantum advantage'. In: Proceedings
  of the IEEE/ACM 42nd International Conference on Software Engineering
  Workshops, pp 427--429

\bibitem[{Brandl(2017)}]{brandl2017quantum}
Brandl MF (2017) A quantum von neumann architecture for large-scale quantum
  computing. arXiv preprint arXiv:170202583

\bibitem[{Businge et~al.(2018)Businge, Openja, Nadi, Bainomugisha, and
  Berger}]{Businge2018CloneBasedVM}
Businge J, Openja M, Nadi S, Bainomugisha E, Berger T (2018) Clone-based
  variability management in the android ecosystem. 2018 IEEE International
  Conference on Software Maintenance and Evolution (ICSME) pp 625--634

\bibitem[{Businge et~al.(2019)Businge, Openja, Kavaler, Bainomugisha, Khomh,
  and Filkov}]{Businge2019StudyingAA}
Businge J, Openja M, Kavaler D, Bainomugisha E, Khomh F, Filkov V (2019)
  Studying android app popularity by cross-linking github and google play
  store. 2019 IEEE 26th International Conference on Software Analysis,
  Evolution and Reengineering (SANER) pp 287--297

\bibitem[{Campos and Souto(2021)}]{Campos2021QBugsAC}
Campos J, Souto A (2021) Qbugs: A collection of reproducible bugs in quantum
  algorithms and a supporting infrastructure to enable controlled quantum
  software testing and debugging experiments. 2021 IEEE/ACM 2nd International
  Workshop on Quantum Software Engineering (Q-SE) pp 28--32

\bibitem[{Catolino et~al.(2019)Catolino, Palomba, Zaidman, and
  Ferrucci}]{CATOLINO2019165}
Catolino G, Palomba F, Zaidman A, Ferrucci F (2019) Not all bugs are the same:
  Understanding, characterizing, and classifying bug types. Journal of Systems
  and Software 152:165--181

\bibitem[{Cramer et~al.(2016)Cramer, Kalb, Rol, Hensen, Blok, Markham,
  Twitchen, Hanson, and Taminiau}]{Cramer}
Cramer J, Kalb N, Rol MA, Hensen B, Blok MS, Markham M, Twitchen DJ, Hanson R,
  Taminiau TH (2016) Repeated quantum error correction on a continuously
  encoded qubit by real-time feedback. Nature communications 7(1):1--7

\bibitem[{Developers(2021)}]{cirq_developers_2021_4586899}
Developers C (2021) Cirq. \doi{10.5281/zenodo.4586899},
  \urlprefix\url{https://doi.org/10.5281/zenodo.4586899}, {See full list of
  authors on Github: https://github .com/quantumlib/Cirq/graphs/contributors}

\bibitem[{Farhi et~al.(2014)Farhi, Goldstone, and Gutmann}]{farhi2014quantum}
Farhi E, Goldstone J, Gutmann S (2014) A quantum approximate optimization
  algorithm. arXiv preprint arXiv:14114028

\bibitem[{Fein(2021)}]{Ecosystem_evo}
Fein R (2021) The evolving quantum computing ecosystem.
  \url{https://medium.com/@russfein/the-evolving-quantum-computing-ecosystem-bed1805007f8},
  accessed: 2022-04-15

\bibitem[{Fingerhuth et~al.(2018)Fingerhuth, Babej, and Wittek}]{Q_2018}
Fingerhuth M, Babej T, Wittek P (2018) Open source software in quantum
  computing. PloS one 13(12):e0208561

\bibitem[{Forn-Diaz(2010)}]{phdthesis}
Forn-Diaz P (2010) Superconducting qubits and quantum resonators. PhD thesis,
  Delft University of Technology

\bibitem[{Foundation(2021)}]{qosf2021opensource}
Foundation QOS (2021) List of open quantum projects.
  \url{https://qosf.org/project\_list/}, {Accessed: 2022-01-10}

\bibitem[{Gao et~al.(2020)Gao, Jin, Zhang, and Ian}]{gao2020pulsequbit}
Gao Y, Jin S, Zhang Y, Ian H (2020) Pulse-qubit interaction in a
  superconducting circuit under frictively dissipative environment. arXiv
  preprint arXiv:200206553

\bibitem[{Garhwal et~al.(2019)Garhwal, Ghorani, and Ahmad}]{garhwal2019quantum}
Garhwal S, Ghorani M, Ahmad A (2019) Quantum programming language: A systematic
  review of research topic and top cited languages. Archives of Computational
  Methods in Engineering pp 1--22

\bibitem[{GitHub(2021)}]{githubrestapi}
GitHub (2021) {REST API}. \url{https://developer.github.com/v3/}, {Last
  accessed 2021-05-04}

\bibitem[{Huang and Martonosi(2019)}]{10.1145/3307650.3322213}
Huang Y, Martonosi M (2019) Statistical assertions for validating patterns and
  finding bugs in quantum programs. In: Proceedings of the 46th International
  Symposium on Computer Architecture, Association for Computing Machinery, New
  York, NY, USA, ISCA '19, p 541–553

\bibitem[{IBM(2021)}]{IBM_q}
IBM (2021) Survey of ibm-q, circuit composer, backends. install qiskit locally
  and get started with tutorials.
  \url{http://www.web.uvic.ca/~rdesousa/teaching/P280/L15\_280.pdf}, accessed:
  2021-10-20

\bibitem[{IonQ(2018)}]{Qsharp}
IonQ (2018) {MS Windows NT} kernel description. \url{https://ionq.com/},
  accessed: 2021-10-20

\bibitem[{Kalliamvakou et~al.(2016)Kalliamvakou, Gousios, Blincoe, Singer,
  Germ{\'a}n, and Damian}]{Kalliamvakou2015AnIS}
Kalliamvakou E, Gousios G, Blincoe K, Singer L, Germ{\'a}n DM, Damian DE (2016)
  An in-depth study of the promises and perils of mining github. Empirical
  Software Engineering 21:2035--2071

\bibitem[{Kaye et~al.(2007)Kaye, Laflamme, Mosca et~al.}]{kaye2007introduction}
Kaye P, Laflamme R, Mosca M, et~al. (2007) An introduction to quantum
  computing. Oxford University Press on Demand

\bibitem[{Knight(2017)}]{IBM_raise}
Knight W (2017) {MS Windows NT} kernel description.
  \url{https://www.technologyreview.com/2017/11/10/147728/ibm-raises-the-bar-with-a-50-qubit-quantum-computer/},
  accessed: 2021-10-20

\bibitem[{LaRose(2019)}]{larose2019overview}
LaRose R (2019) Overview and comparison of gate level quantum software
  platforms. Quantum 3:130

\bibitem[{Li et~al.(2020)Li, Zhou, Yu, Ding, Ying, and Xie}]{10.1145/3428218}
Li G, Zhou L, Yu N, Ding Y, Ying M, Xie Y (2020) Projection-based runtime
  assertions for testing and debugging quantum programs. Proceedings of the ACM
  on Programming Languages 4(OOPSLA):1--29

\bibitem[{Moguel et~al.(2020)Moguel, Berrocal, Garc{\'\i}a-Alonso, and
  Murillo}]{moguel2020roadmap}
Moguel E, Berrocal J, Garc{\'\i}a-Alonso J, Murillo JM (2020) A roadmap for
  quantum software engineering: Applying the lessons learned from the classics.
  In: Q-SET@ QCE, pp 5--13

\bibitem[{Mueck(2017)}]{mueck2017quantum}
Mueck L (2017) Quantum software. Nature 549(171)

\bibitem[{Nachman et~al.(2020)Nachman, Urbanek, de~Jong, and
  Bauer}]{nachman2020unfolding}
Nachman B, Urbanek M, de~Jong WA, Bauer CW (2020) Unfolding quantum computer
  readout noise. npj Quantum Information 6(1):1--7

\bibitem[{Paltenghi and Pradel(2021)}]{abs-2110-14560}
Paltenghi M, Pradel M (2021) Bugs in quantum computing platforms: An empirical
  study. CoRR abs/2110.14560

\bibitem[{Physicsworld.com(2011{\natexlab{a}})}]{DWave2011}
Physicsworldcom (2011{\natexlab{a}}) Quantum-computing firm opens the box.
  \url{https://web.archive.org/web/20110515083848/http://physicsworld.com/
  cws/article/news/45960}, {Accessed: 2022-01-10}

\bibitem[{Physicsworld.com(2011{\natexlab{b}})}]{IonQ}
Physicsworldcom (2011{\natexlab{b}}) Quantum-computing firm opens the box.
  \url{https://physicsworld.com/a/ion-based-commercial-quantum-computer-is-a-first/},
  {Accessed: 2022-01-10}

\bibitem[{Piattini et~al.(2020{\natexlab{a}})Piattini, Peterssen, and
  P{\'e}rez-Castillo}]{piattini2020quantum}
Piattini M, Peterssen G, P{\'e}rez-Castillo R (2020{\natexlab{a}}) Quantum
  computing: A new software engineering golden age. ACM SIGSOFT Software
  Engineering Notes 45(3):12--14

\bibitem[{Piattini et~al.(2020{\natexlab{b}})Piattini, Peterssen,
  P{\'e}rez-Castillo, Hevia, Serrano, Hern{\'a}ndez, de~Guzm{\'a}n, Paradela,
  Polo, Murina et~al.}]{piattini2020talavera}
Piattini M, Peterssen G, P{\'e}rez-Castillo R, Hevia JL, Serrano MA,
  Hern{\'a}ndez G, de~Guzm{\'a}n IGR, Paradela CA, Polo M, Murina E, et~al.
  (2020{\natexlab{b}}) The talavera manifesto for quantum software engineering
  and programming. In: The First International Workshop on the Quantum Software
  Engineering \& Programming, QANSWER '20, pp 1--5

\bibitem[{Piattini et~al.(2021)Piattini, Serrano, Perez-Castillo, Petersen, and
  Hevia}]{piattini2021toward}
Piattini M, Serrano M, Perez-Castillo R, Petersen G, Hevia JL (2021) Toward a
  quantum software engineering. IT Professional 23(1):62--66

\bibitem[{{Replication Package}(2022)}]{ReplicationPackage}
{Replication Package} (2022) Study of the bug charateristics in the quantum
  ecosystem. \url{https://github.com/raed1337/Quantum-bug-study/tree/master},
  accessed: 2022-04-20

\bibitem[{Roy and Devoret(2018)}]{Roy_2018}
Roy A, Devoret M (2018) Quantum-limited parametric amplification with josephson
  circuits in the regime of pump depletion. Physical Review B 98(4):045405

\bibitem[{Sodhi(2018)}]{Sodhi2018QualityAO}
Sodhi B (2018) Quality attributes on quantum computing platforms. arXiv
  preprint arXiv:180307407

\bibitem[{Tan et~al.(2013)Tan, Liu, Li, Wang, Zhou, and Zhai}]{Tan2013BugCI}
Tan L, Liu C, Li Z, Wang X, Zhou Y, Zhai C (2013) Bug characteristics in open
  source software. Empirical Software Engineering 19:1665--1705

\bibitem[{TechTarget(2018)}]{ClassicalComputer}
TechTarget (2018) classical computing. \url{https://www.techtarget.com/
  whatis/definition/classical-computing}, {Accessed: 2022-04-18}

\bibitem[{Wootters and Zurek(1982)}]{Wootters1982ASQ}
Wootters WK, Zurek W (1982) A single quantum cannot be cloned. Nature
  299:802--803

\bibitem[{Yu and Palsberg(2021)}]{10.1145/3453483.3454061}
Yu N, Palsberg J (2021) Quantum abstract interpretation. In: Association for
  Computing Machinery, Association for Computing Machinery, New York, NY, USA,
  PLDI 2021, p 542–558

\bibitem[{Zhao(2020)}]{zhao2020quantum}
Zhao J (2020) Quantum software engineering: Landscapes and horizons. arXiv
  preprint arXiv:200707047

\bibitem[{Zhao et~al.(2021)Zhao, Zhao, and Ma}]{abs-2103-09069}
Zhao P, Zhao J, Ma L (2021) Identifying bug patterns in quantum programs. In:
  2021 IEEE/ACM 2nd International Workshop on Quantum Software Engineering
  (Q-SE), IEEE, pp 16--21

\end{thebibliography}

\end{document}